\documentclass[%
 aip,
 amsmath,amssymb,
 reprint,
]{revtex4-1}
\usepackage[utf8]{inputenc}
\usepackage{graphicx}
\usepackage[version=4]{mhchem}
\usepackage{bm}
\usepackage{cases}
\usepackage{upgreek}
\usepackage[usenames, dvipsnames]{color}
\usepackage[normalem]{ulem}
\usepackage[frozencache,cachedir=minted-cache]{minted}
\usepackage{hyperref}

\makeatletter
\def\@email#1#2{%
 \endgroup
 \patchcmd{\titleblock@produce}
  {\frontmatter@RRAPformat}
  {\frontmatter@RRAPformat{\produce@RRAP{*#1\href{mailto:#2}{#2}}}\frontmatter@RRAPformat}
  {}{}
}%
\makeatother
\begin{document}
\preprint{AIP/123-QED}

\title{Trion formation resolves observed peak shifts in the optical spectra of transition metal dichalcogenides
\vspace{8pt}}

\author{Thomas Sayer}
\affiliation{Department of Chemistry, University of Colorado Boulder; Boulder, CO, USA}

\author{Yusef R. Farah}
\affiliation{Department of Chemistry, Colorado State University; Fort Collins, CO, USA}

\author{Rachelle Austin}
\affiliation{Department of Chemistry, Colorado State University; Fort Collins, CO, USA}

\author{Justin Sambur}
\homepage{Justin.Sambur@colostate.edu}
\affiliation{Department of Chemistry, Colorado State University; Fort Collins, CO, USA}
\affiliation{School of Advanced Materials Discovery, Colorado State University; Fort Collins, CO, USA\looseness=-1}

\author{Amber T. Krummel}
\homepage{Amber.Krummel@colostate.edu}
\affiliation{Department of Chemistry, Colorado State University; Fort Collins, CO, USA}

\author{\\Andr\'{e}s Montoya-Castillo}
\homepage{Andres.MontoyaCastillo@colorado.edu}
\affiliation{Department of Chemistry, University of Colorado Boulder; Boulder, CO, USA} 


\date{14 March 2023}

\begin{abstract}
\vspace{5pt}
Monolayer transition metal dichalcogenides (TMDs) have the potential to unlock novel photonic and chemical technologies if their optoelectronic properties can be understood and controlled. Yet, recent work has offered contradictory explanations for how TMD absorption spectra change with carrier concentration, fluence, and time. Here, we test our hypothesis that the large broadening and shifting of the strong band-edge features observed in optical spectra arise from the formation of negative trions. We do this by fitting an \textit{ab initio}-based, many-body model to our experimental electrochemical data. Our approach provides an excellent, global description of the potential-dependent linear absorption data. We further leverage our model to demonstrate that trion formation explains the non-monotonic potential dependence of the transient absorption spectra, including through photoinduced derivative line shapes for the trion peak. Our results motivate the continued development of theoretical methods to describe cutting-edge experiments in a physically transparent way.
\vspace{20pt}
\end{abstract}

\maketitle

Monolayer transition metal dichalcogenides (ML-TMDs) are promising lightweight and flexible electrode materials for solid-state and electrochemical devices. For light-driven applications such as photovoltaics and photodetectors, optimizing device efficiency requires a fundamental understanding of optoelectronic properties.

Optical spectroscopies of TMDs reveal exotic phenomena, including significant band gap renormalization (BGR) and screening of their quasiparticle binding energies (BES). These two effects determine the energy of the band-edge exciton and tunes its shift as a function of material parameters, such as: TMD composition, substrate, superstrate (often liquid electrolyte in an electrochemical cell), and doping density (here, the presence of electrons in the conduction band).\cite{Berkelbach2018b} For example, the \textit{ab initio} GW~value for the electronic band gap of unsupported ML-\ce{MoS2} is 2.8~eV,\cite{Qiu2013,Ryou2016,Ataei2021,Austin2022}$^,$\footnote{More converged calculations reduce the GW~value to 2.64~eV.\cite{Qiu2015}} whereas its measured band gap is 2.17~eV on graphite\cite{Zhang2014a} and 1.39~eV on gold\cite{Miwa2015}. Crucially, the screening responsible for BGR concomitantly weakens the attraction between electrons and holes in the band-edge excitons. This BES is a countervailing effect which opposes the BGR shift in the optical signal. Hence, only small changes in the optical gap are experimentally observed.\cite{Ugeda2014,Gies2021,Austin2022}

TMDs also exhibit other many-body effects that pose a challenge to their first principles exploration and obfuscate the interpretation of optical measurements. In particular, exciton-electron quasiparticles (negative trions) have been observed in \ce{WS2},\cite{Liu2016b, Zhao2020c, Zipfel2020a, Lan2020, Muir2022}, \ce{WSe2},\cite{Courtade2017,Li2018c,Wagner2020,Lan2020,Yang2022a} \ce{MoSe2},\cite{Li2021c} and \ce{MoS2}\cite{Mak2012, Liu2016b, Kim2019, Huang2022}. Fundamentally, the extent to which trion formation affects the behavior of working devices at room temperature remains unclear.\cite{Lui2014, Goldstein2020a} This is of widespread concern because TMDs are incidentally doped by defects which form during synthesis\cite{Zafar2017,xyzValence2019}, and can also acquire charge from electrolyte solutions\cite{Lin2014}: the resulting conduction band electrons induce trions to form, which causes oscillator strength to transfer from the exciton to the trion signal in optical experiments.\cite{Chernikov2015} A model of minimal complexity that can predict how variations in the microscopic parameters translate to changes in measured quantities, such as spectral features, is therefore essential to disentangle the role of trions and achieve predictive power.

\begin{figure}[!h]
\vspace{-8pt}
  \centering
  \begin{minipage}[c]{0.29\textwidth}
  \vspace{4pt}
    \resizebox{1.0\textwidth}{!}{\includegraphics{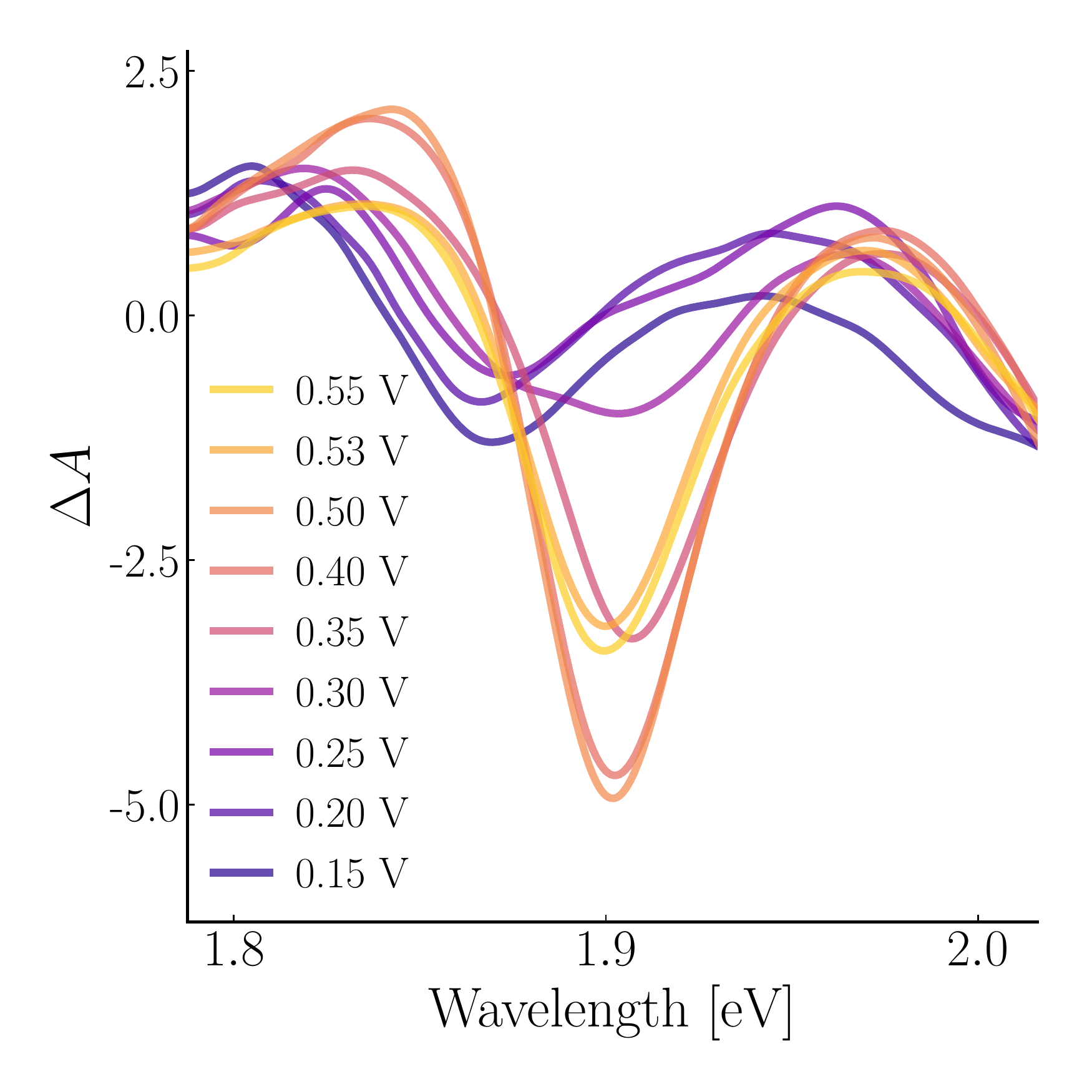}}
  \end{minipage}
  \begin{minipage}[c]{0.18\textwidth}
    \resizebox{1.0\textwidth}{!}{\includegraphics[trim={2.8cm 9.7cm 8.1cm 4cm},clip]{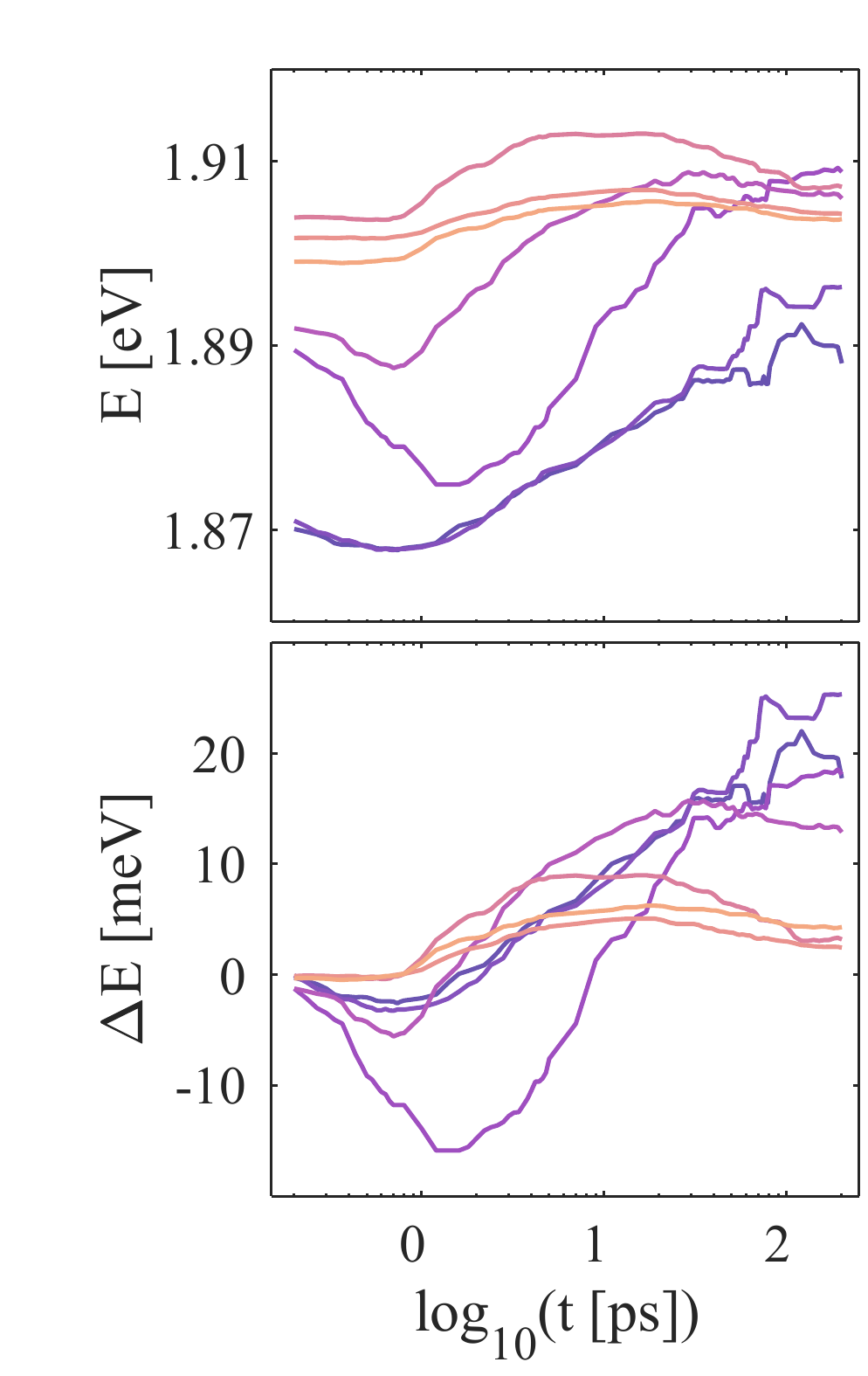}}
  \end{minipage}
  \vspace{-6pt}
  \caption{\label{fig:initial_spectra} Dynamic peak shifts and intensity changes of a ML-\ce{MoS2} electrode in a working electrochemical cell. \textbf{Left}: Potential-dependent TA spectra of the A-exciton region at 1~ps delay time after a 3.1~eV pump pulse. Appendix~Fig.~\ref{fig:sibar1} shows the full range of the data set. All spectra were acquired in 1~M NaI electrolyte versus a Ag/AgI reference electrode and a Pt counter electrode. \textbf{Right}: Plot of the ``superpeak'' position (most negative value) in the TA against time in terms of (\textbf{top}) absolute and (\textbf{bottom}) relative shift. A sub-picosecond redshift in some potentials is followed by a net blueshift, which is most pronounced at low potentials.}
  \vspace{0pt}
\end{figure}

These challenges of interpretation have become prominent in recent years, as experiments have revealed that ML-TMDs display large, complex spectral shifts in device architectures such as photoelectrochemical cells,\cite{Carroll2019} field effect transistors,\cite{Chen2019e, Pradeepa2020} and photovoltaics\cite{Chen2019f}. For example, Sie \textit{et al}.~invoked photoexcitation-induced changes in BGR and BES to explain the spectral shifts observed when pumping ML-\ce{WS2} with increasingly higher laser fluence.\cite{Sie2017} However, since the observations were inconsistent with this explanation, the authors posited a more sophisticated model based on exciton-exciton interactions. More recently, we have reported hot-carrier extraction from a ML-$\rm{MoS}_2$ photoelectrode in a working photoelectrochemical cell which displayed complex changes in the spectral features as a function of applied potential.\cite{Austin2022} As increasingly negative potentials raise the conduction band population, the applied potential affects the magnitudes of BGR and BES. Indeed, we observed a large potential dependence of the position, height, and width of the band-edge ``A-exciton'' peak, analogous to Ref.~\onlinecite{Sie2017}. The potential-dependence at a given time, or `static shift', is evident in the experimental transient absorption (TA)~spectra at 1~ps displayed in Fig.~\ref{fig:initial_spectra}. As the photocarriers relax back to their ground state, the peaks' heights and positions undergo both red and blue `dynamic shift' in time (Fig.~\ref{fig:initial_spectra}(right)). These observations are significant because the mechanism of charge transfer and the role of excitons in these devices are not yet fully understood.\cite{Hong2014,Zimmermann2021,Kim2019,Johansson2020,Xu2021} The difficulty of interpreting these shifts illustrates the need for utilizable and transparent frameworks that facilitate interpretation of experimental data. 

Here, we provide a unifying framework based on the doping-dependent emergence of trions to explain the complex behavior of both potential-\cite{Austin2022} and fluence-dependent\cite{Sie2017, Bera2021} spectra. Since in our experiment both the laser pulse and the applied potential affect BGR and BES, the observation of a qualitatively equivalent behavior\footnote{This study further demonstrated a similar time-dependent shift of the band-edge exciton position over the period of electronic relaxation (see Fig.~4 in Ref.~\onlinecite{Sie2017}).} in a different TMD under entirely different experimental conditions suggests the behavior is a signature of the same underlying photophysical process. Indeed, another recent experiment focusing on $\rm{MoS}_2$ further established the universality of this shift in TMDs.\cite{Bera2021} In our analysis, we make two key observations. First, careful inspection of both sets of spectra suggest that they are not always well described by a single peak. Second, recent advances in GW that include free carrier screening predict that the \ce{WS2} A-exciton should neither redshift nor blueshift with increasing carrier density, and---more strikingly---that the absorption spectrum of $\rm{MoS}_2$ should undergo a pronounced and monotonic blueshift with increasing carrier density.\cite{Ataei2021} Both sets of spectra disagree with the cutting-edge theoretical prediction considering only BGR and BES. We resolve this discrepancy by demonstrating that the source of both static and dynamical peak shifts is the formation of the trion, which is the next-most significant many-body effect in these systems.

\pagebreak

\begin{figure}[h]
\vspace{10pt}
\begin{center} 
    \resizebox{.31\textwidth}{!}{\includegraphics{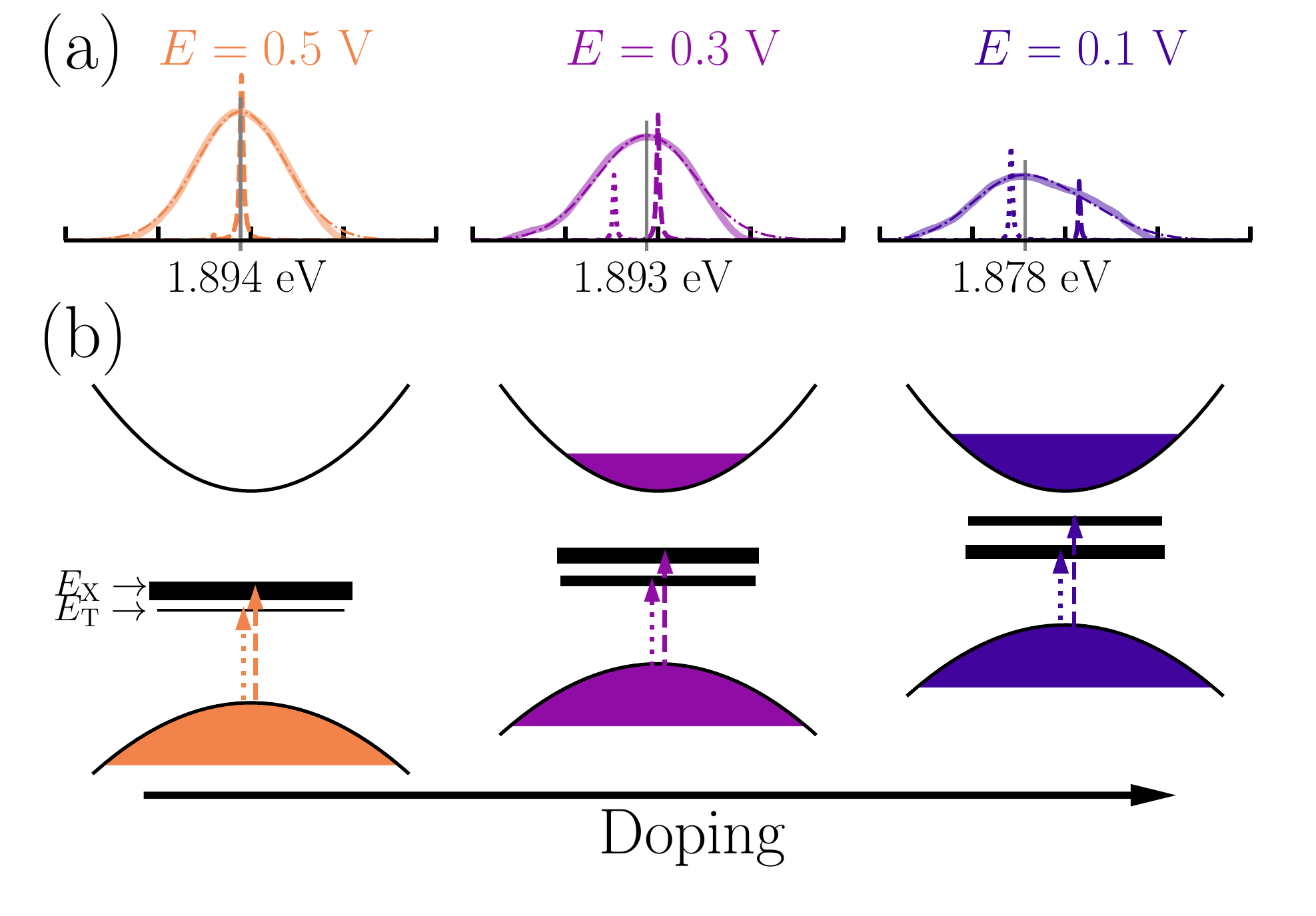}}
    \resizebox{.31\textwidth}{!}{\includegraphics{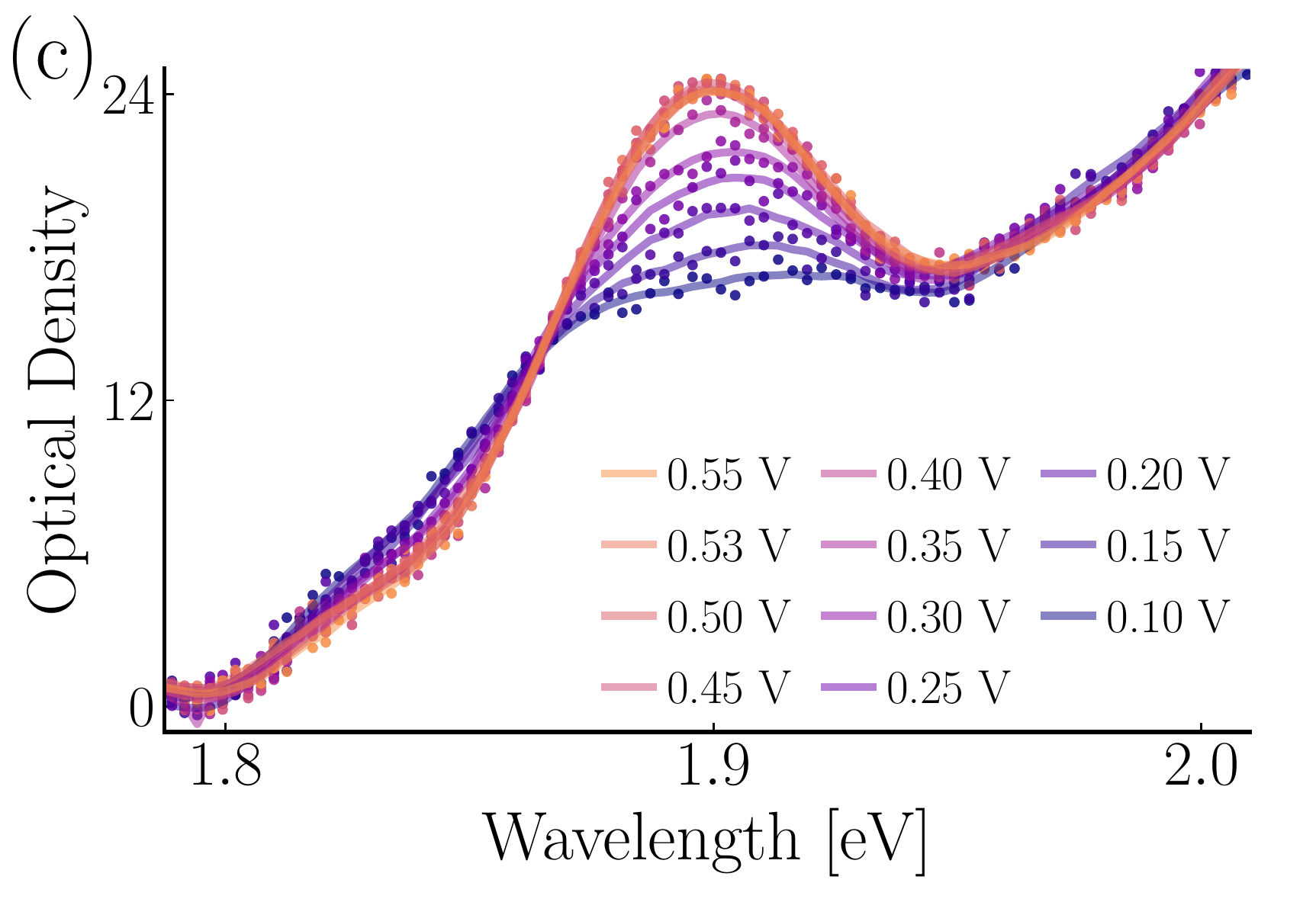}}  
    \resizebox{.31\textwidth}{!}{\includegraphics{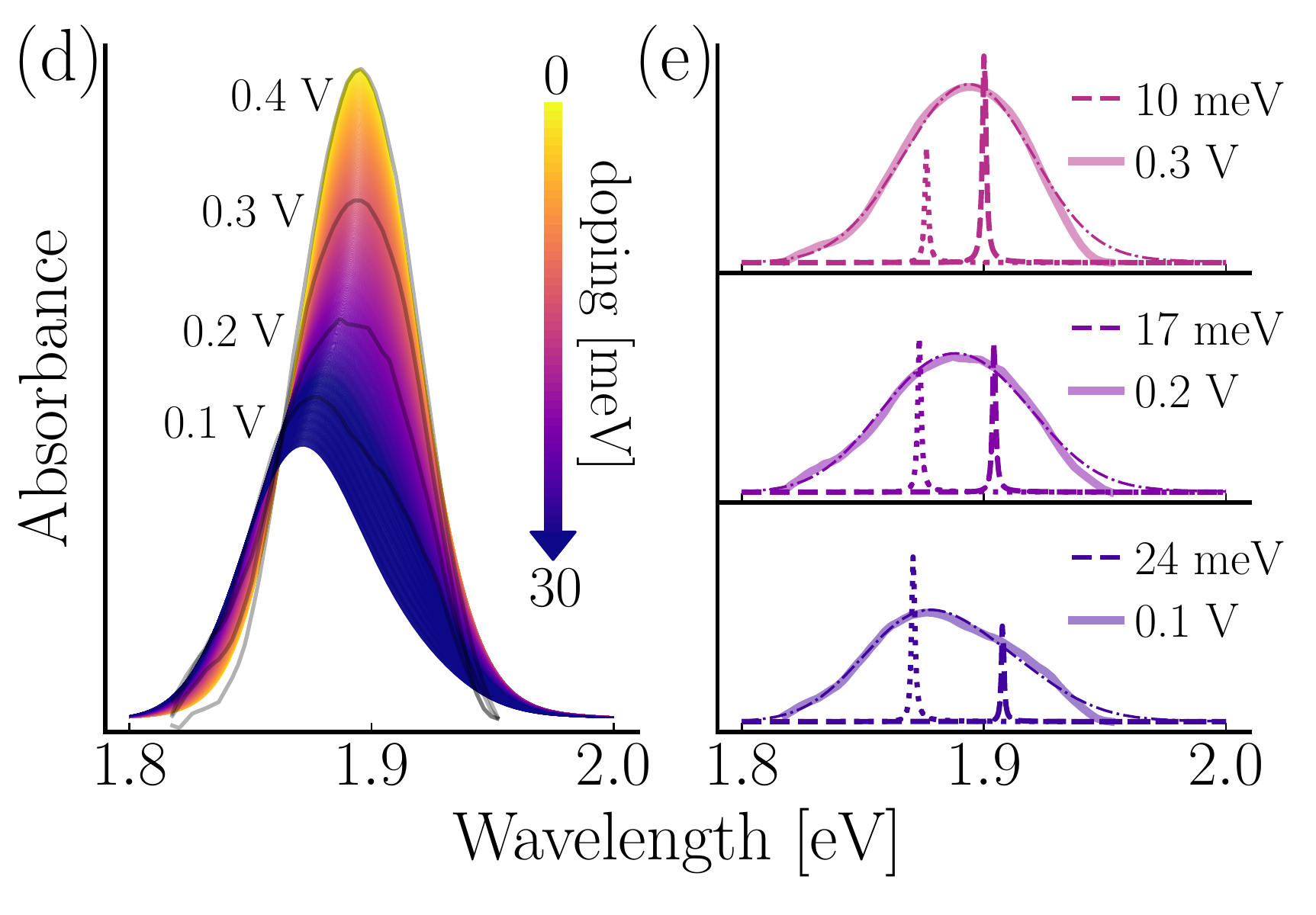}}
\end{center}
\vspace{-16pt}
\caption{\label{fig:steadystate} Summary of the quantitative analysis procedure to model potential-dependent linear absorption spectra with the MND model. \textbf{(a)}~Comparison of theory and simulation for low (0.5~V), medium (0.3~V), and high (0.1~V) doping conditions. Narrow dashed and dotted lines represent exciton and trion absorption peaks, respectively. The broad dash-dotted and solid lines represent the convoluted simulation result and experimental data, respectively. \textbf{(b)}~Schematic of the MND model. More doping corresponds to: a higher Fermi level and a lower applied voltage; oscillator strength shifting to the trion, and the binding energy of the trion (difference between the mid-gap levels) increasing. The BGR occurs exclusively in the valence band.\cite{xyzValence2019} \textbf{(c)}~Potential-dependent experimental linear spectra (points) and the minimal smooths obtained via SavGol filter (solid line). \textbf{(d)}~Manifold of curves for a given set of MND parameters (peak heights, binding energies, line widths) allowed over the doping range 0--30~meV. \textbf{(e)}~The linear spectra for 4~representative voltages are superimposed as fin gray lines, and a close-up of each of the three trion-containing peaks is displayed to the side. The model is parameterised to optimize the fit to all curves simultaneously.}
\end{figure}

While a number of descriptions of the trion are available\cite{Chang2018, Chang2019a, Torche2019, Zhumagulov2020, Rana2021, Glazov2020b, Katsch2022}, we adopt the Mahan-Nozi{\'e}res-De Dominicis (MND) Hamiltonian\cite{Mahan2000}---a minimal many-body model of a Fermi polaron consisting of a heavy exciton dressed by a free electron bath---which prioritizes the description of doping-dependent phenomena.\cite{Chang2019a} This approach describes how changing the conduction band occupation causes the trion and exciton peaks to shift in energy, change line shape, and transfer oscillator strength between one another, as presented in Fig.~\ref{fig:steadystate}(a). We also note that, like Ref.~\onlinecite{Ataei2021}, it predicts the A-exciton peak to blueshift with increasing doping.
\onecolumngrid

\twocolumngrid

In the following, we interrogate the ability of this framework to capture the features that arise both in our experimental setup\cite{Austin2022} and in the fluence-dependent studies of Refs.~\onlinecite{Sie2017, Bera2021}. As this theory is designed to describe linear spectra, we first ratify its use by capturing the potential-dependent linear absorption spectra shown in Fig.~\ref{fig:steadystate}(c). At this \ce{ITO_{(s)}|MoS2_{(ML)}|I- / I3^-_{(aq)}} interface the band-edge peak moves to higher energy with more positive applied potential. We claim that the two peaks predicted by the MND theory should be convolved with a Gaussian representing the combined effect of broadening mechanisms and the instrument response to obtain the observed ``superpeak''. That is, when the doping increases: the trion redshifts, the exciton blueshifts, both peaks flatten slightly,\cite{Carbone2020b} and the oscillator strength transfers to the trion peak, as sketched in Fig.~\ref{fig:steadystate}(b). Then, the phenomenological broadening of this total signal will reproduce the shift and change in line shape observed in the experiments.

\begin{figure}[!b]
\vspace{-12pt}
\begin{center}
    \resizebox{.45\textwidth}{!}{\includegraphics{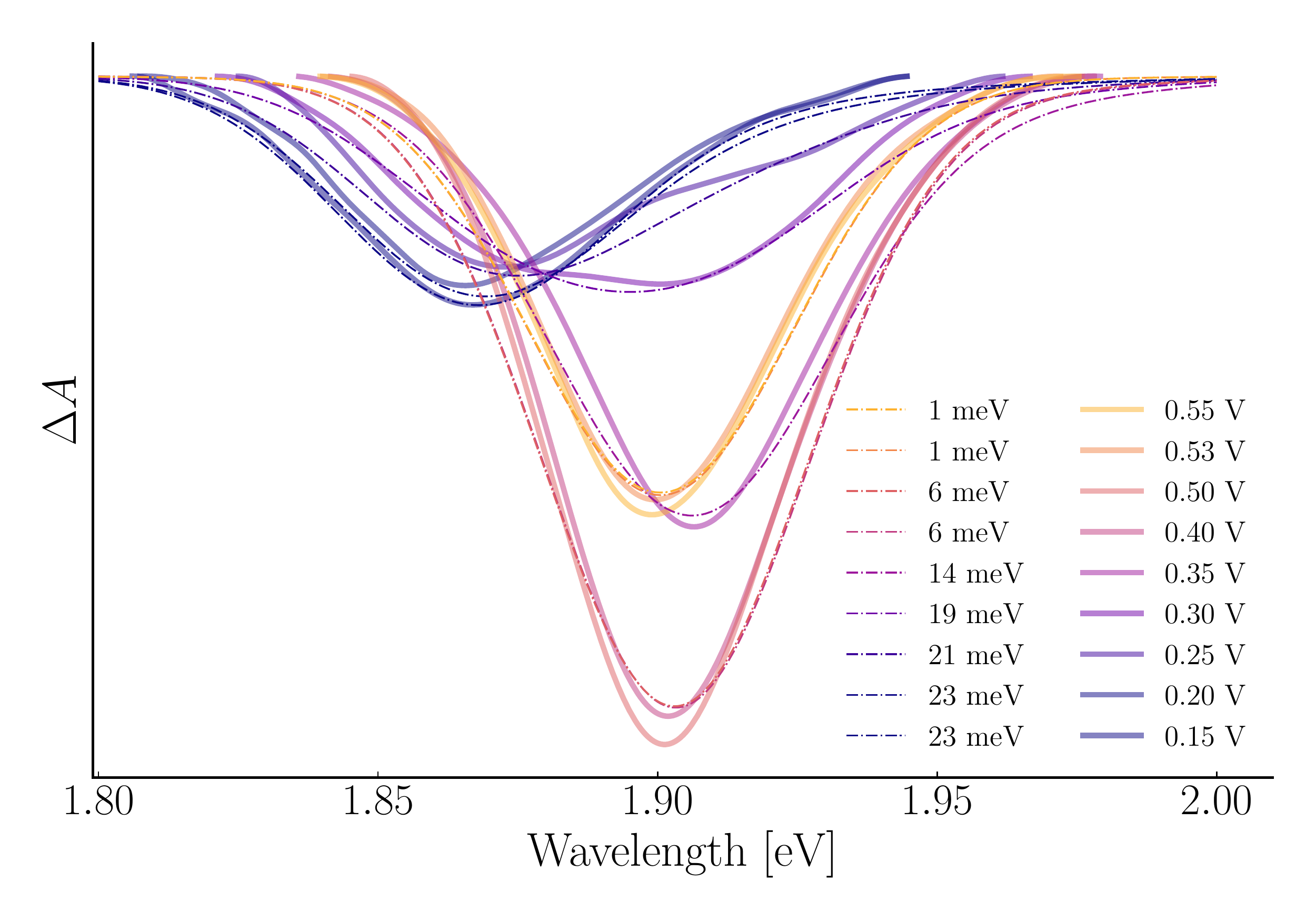}}
\end{center}
\vspace{-18pt}
\caption{\label{fig:TAfit1ps} Comparison of MND theory to the TA spectrum at 1~ps shown in Fig.~\ref{fig:initial_spectra}. The TA spectra are shown as full lines, and the MND prediction is dash-dotted. Each TA curve has had its linear background removed, been cropped at the positions of the photoinduced maxima either side of the bleach signal and baselined (see Appendix~\ref{app:preprocess}). Note: as the model is now parameterized to TA~data, the doping values 0–-30~meV no longer correspond precisely to the original values of Ref.~\onlinecite{Chang2019a}, and should be considered as some arbitrary, internal units.}
\vspace{-4pt}
\end{figure}

Use of the MND description requires fitting to the experimental data because we are currently unable to predict certain quantities from \textit{ab initio} calculations. For example, Ref.~\onlinecite{Chang2019a} assumed that the $\rm{MoS}_2$~ML is a defect-free, intrinsic semiconductor at 0~K and unsupported \textit{in vacuo}. In our experimental cell, the balance of BGR and BES can be expected to differ. We will work with the theory not by optimizing for these Hamiltonian parameters by repeatedly solving a many-body equation, but rather by modelling it at the level of its outputs: peak positions, heights, and widths as functions of doping. This is convenient because these functions are predicted to have a simple (near)-linear form,\cite{Chang2019a} such that the gradients and intercepts of these functions become the only free parameters. While this fitting offers no guarantee that a given set of functions all correspond to the same (or any) Hamiltonian, it serves as an accessible starting point for interpreting multifaceted experimental data sets. We describe our protocol for how to extract the model parameters from the potential-dependent photoelectrochemical spectra, which matches the spectra with a manifold of spectral curves as a function of applied potential (e.g. Fig.~\ref{fig:steadystate}(d)), in Appendix~\ref{app:optimization-procedure}.  

Figure~\ref{fig:steadystate}(e) shows the result for our linear absorption spectra, which demonstrates the model provides an excellent global fit to the data. As Figs.~\ref{fig:steadystate}(a)~and~(e) illustrate, the static shift of the superpeak with applied potential is mostly due to the transfer of oscillator strength between exciton and trion, with a smaller contribution from the shifting of the underlying peaks. As the potential becomes more positive, the superpeak shifts from the trion position to the exciton position. 

Can this simple explanation for the static shift shed light on the source of the more complex dynamical shift presented in Fig.~\ref{fig:initial_spectra}? After all, TA~is a difference-method where the signal arises from the interplay of the pump and probe pulses, with $\Delta A(\tau) = A(\tau) - A_\mathrm{SS}$, where $A(\tau)$ is the absorption of the excited material, $\tau$ units after the pump, and $A_\mathrm{SS}$ is the unpumped steady-state spectrum. 

To simplify the analysis, we discard the TA~data before $1$~ps, where there is a qualitatively different regime which has been assigned to relaxed free carriers at the band-edge waiting to form excitons.\cite{Ceballos2016, Cunningham2017} However, regardless of the time chosen, we must baseline correct the TA~spectra to isolate the peaks for modelling (see Appendix~\ref{app:preprocess}). This introduces a complication since the initial pump pulse, which generates carriers in both the valence and conduction bands, alters the BGR and BES contributions seen by the probe pulse. The resulting shift of the TA~peaks due to this renormalization results in photoinduced absorption features that appear either side of the main bleach signal,\cite{Pogna2016,Carroll2019,Kastl2022} which is a general effect that arises in the TA~of semiconductors.\cite{Price2015} For us, the trion and exciton peaks contaminate each other in a non-trivial, doping-dependent way. Yet, as we show in the following, our approach allows us to discern the emergence of the trion peak in TA~data, its interplay with the exciton peak, and how these peaks subtly shift in TA~experiments. 

Figure~\ref{fig:TAfit1ps} demonstrates that our approach offers a compelling fit to the TA data at $1$~ps, reproducing the observed shifts, peak splittings, and height changes. Given the minimal, but physics-based, nature of the model, some disagreement may be expected. After all, our baseline correction misses the transient effects of photoinduced BGR. Moreover, even if the linear spectra are well described by this model, there is no guarantee the same holds in the TA. In fact, one can readily observe (even in Fig.~\ref{fig:initial_spectra}(left) before the processing) that the height of the superpeak unexpectedly increases as the potential is lowered from 0.55~V to 0.4~V, before it falls and shifts, as previously observed in the linear spectra. While this might indicate that the MND theory is not applicable here, the fit of Fig.~\ref{fig:TAfit1ps} captures the trend near-quantitatively. 

How can the model accurately capture these TA-specific changes in peak height? The linear spectra are subject to a sum rule where the total oscillator strength is a conserved quantity, and the relative share between trion and exciton is all that changes with carrier density. Our algorithm does not incorporate this rule because TA~does not follow a similar requirement. Inspecting the results of the fit for the $1$~ps data of Fig.~\ref{fig:TAfit1ps} shows that making the potential more negative causes the trion intensity to rise more than it causes the exciton intensity to fall. When combined with the small binding energy at low doping (the peaks begin close together), this allows for the initial growth of the superpeak not observed in the linear spectra. Thus, we can assign this effect to the breaking of the sum rule. 

\begin{figure}[!t]
\begin{center}
    \resizebox{.29\textwidth}{!}{\includegraphics{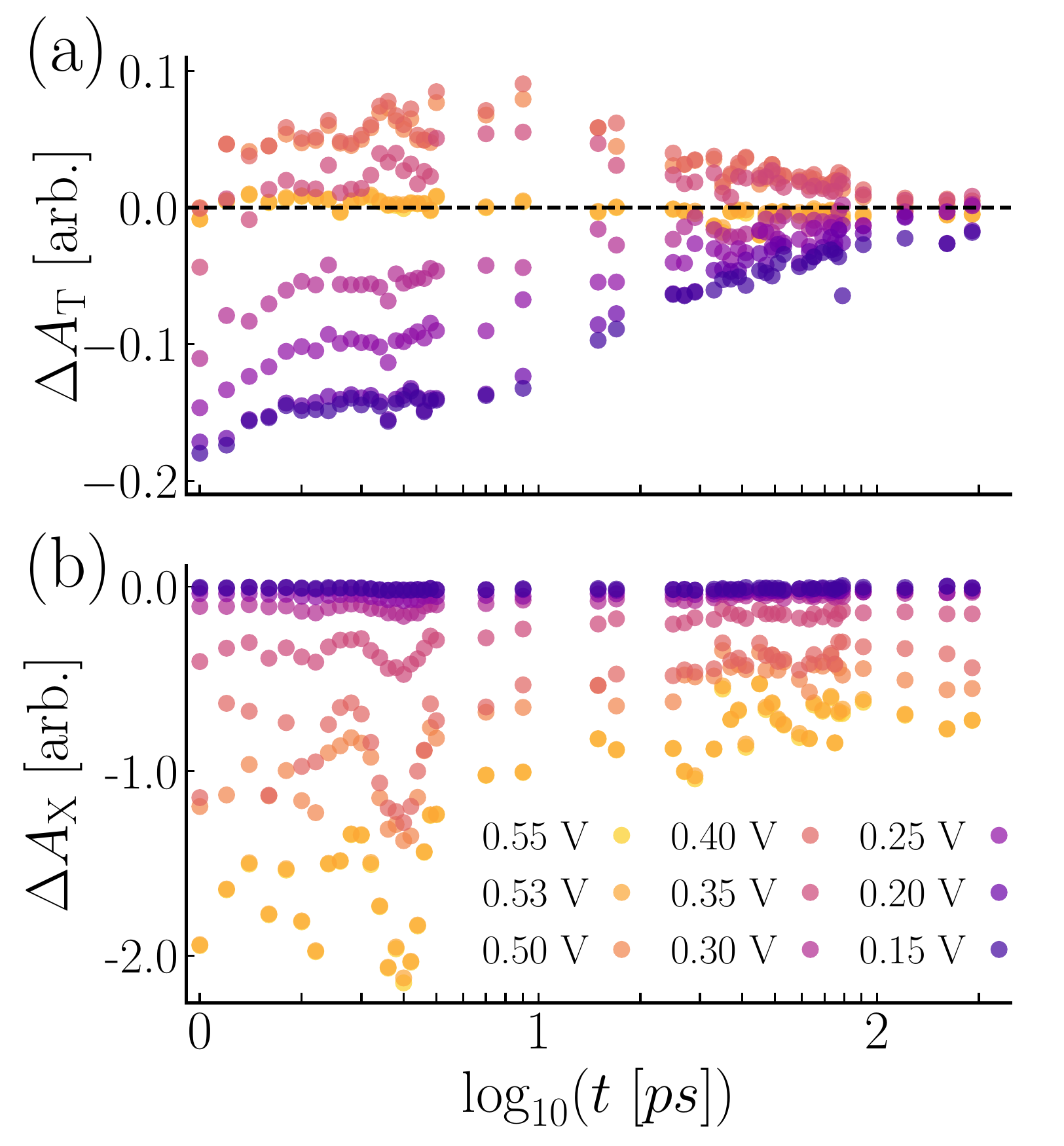}}
    \resizebox{.18\textwidth}{!}{\includegraphics{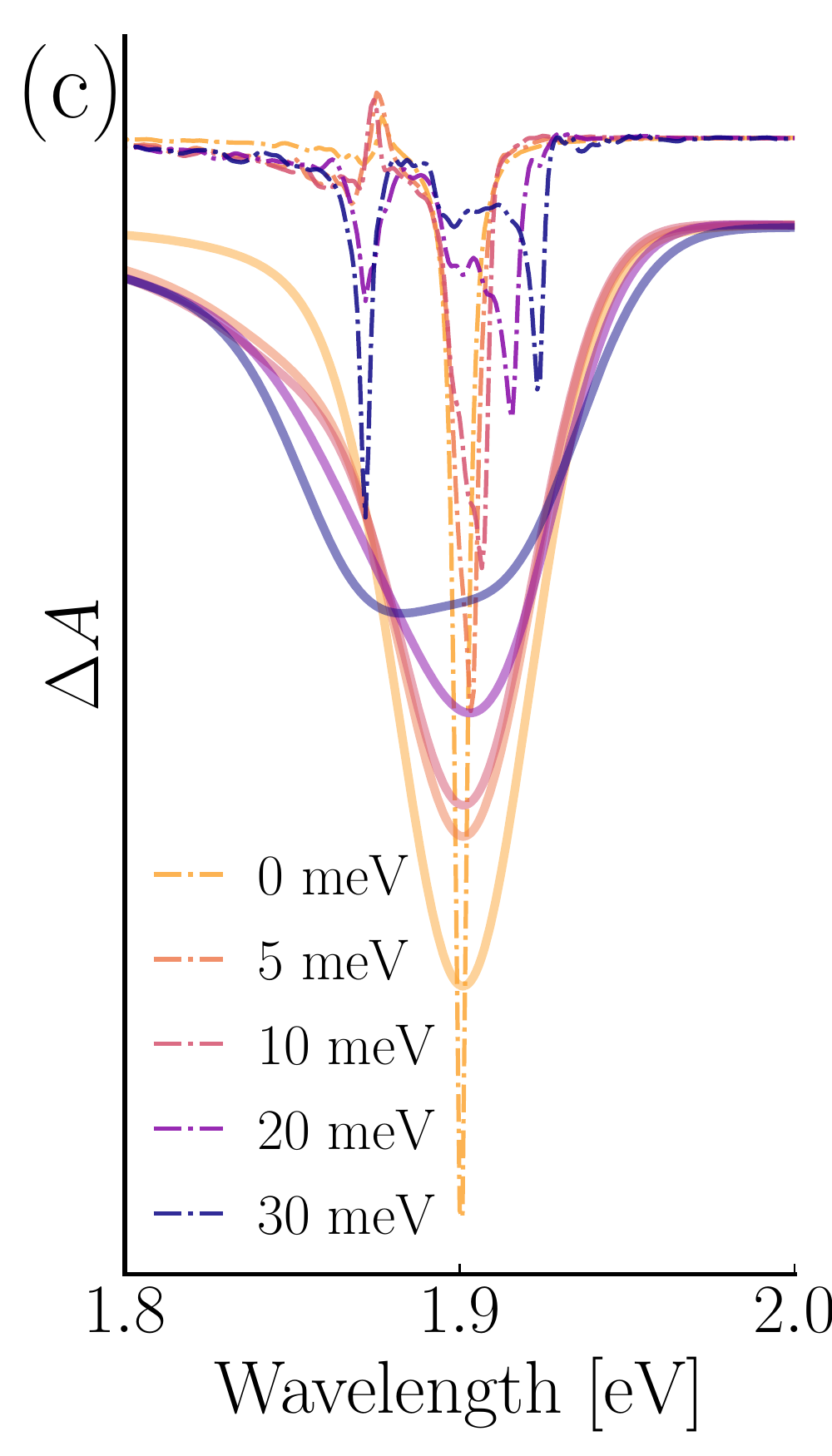}}
\end{center}
\vspace{-14pt}
\caption{\label{fig:TAresults} \textbf{Left}: Plots showing how the repeated fitting of the MND model to the TA~data over time, this gives the equivalent of Fig.~\ref{fig:TAfit1ps} at each time point. Gaps in the series are due to experimental anomalies described in Appendix~\ref{sec:TAresults}. \textbf{(a)} Contribution at each voltage due to the trion peak. Note how the trion is assigned positive values at intermediary voltages (low, but non-zero doping). \textbf{(b)} Contribution at each voltage due to the exciton peak. The `hill’-like feature around 4~ps is a signature of the model straining to allow for the positive trion signal.  \textbf{Right}: \textbf{(c)} Projection-slice TA spectrum from the 2D response function code of Ref.~\onlinecite{Lindoy2022}. At low dopings, the derivative shape of the trion is seen as a partly-positive feature for potentials up to 10~meV. Solid curves are the result of convolving with a Gaussian. The doping levels are appropriate for the original model Ref.~\onlinecite{Chang2019a} and not the post-optimization Fig.~\ref{fig:TAfit1ps}.}
\end{figure}

We now run the fit at each time point over our $>200$~ps of TA data, using the results at the previous time point as the initial guess. This yields a surprising result: the trion appears with positive $\Delta A$ at low doping (see Fig.~\ref{fig:TAresults}(a)). This assignment would be impossible in linear absorption, but contemporaneous theoretical developments suggest that our fitting algorithm is actually yielding physically correct results. To show this we combined the 2D electronic spectroscopy of the MND Hamiltonian\cite{Lindoy2022} with the projection-slice theorem\cite{Hamm2011} to simulate the (early time) TA spectra of the ML-MoS$_2$ MND Hamiltonian. Figure~\ref{fig:TAresults}(c) shows the result of these calculations. While the nonlinear spectroscopic treatment is qualitatively similar to the linear case, the trion exhibits a derivative line shape at low doping that results in positive intensity on the lower energy side of the exciton peak. This outlines the dangers of deconvolving TA~spectra by fitting positive Gaussian peaks, as the pump pulse can brighten and shift the trion peak compared to the linear spectrum, as well as induce interference with the exciton.\cite{Tempelaar2019}

We return to our initial query concerning the dynamical shift of the superpeaks in the TA spectra first presented in Fig.~\ref{fig:initial_spectra}(right). As Fig.~\ref{fig:TAresults}(a)~and~(b) illustrate, the exciton and trion peaks both lose intensity over time. This is reasonable, as we expect the relaxation of photoexcited species to lower the TA~bleach signal. We offer further confirmation of this in Appendix Fig.~\ref{fig:si1}, which shows that the model's assignment of the doping level is generally stable over time. Moreover, the time series also reveal that both the trion and exciton intensities decay exponentially, with the trion feature decaying faster than the exciton (Appendix Fig.~\ref{fig:si2}). This behavior can arise if trions and excitons decay slower than they interconvert: the relaxation or trapping of excitons causes the trion population to collapse as the equilibrium moves to stabilize the ratio of each species.\cite{Huang2022} Such dynamical equilibrium would imply that, as the material relaxes, the superpeak shifts from a position between the exciton and trion signals to being almost entirely atop the exciton peak. 

\begin{figure}[!t]
\vspace{-12pt}
\begin{center}
    \resizebox{.4\textwidth}{!}{\includegraphics{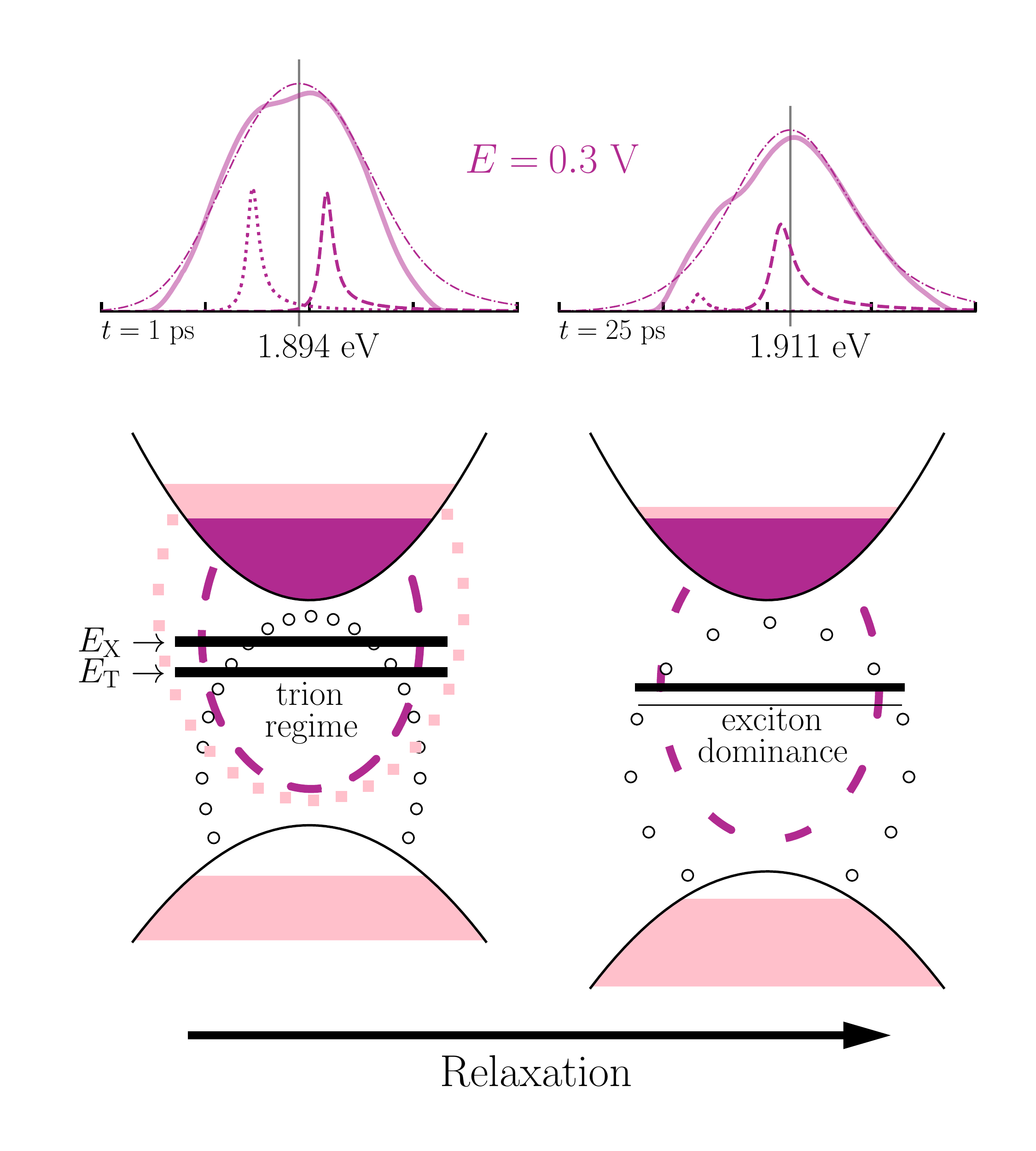}}
\end{center}
\vspace{-24pt}
\caption{\label{fig:explanation} Cartoon illustration showing the relationship between bandgap, carrier concentration, and exciton/trion absorption peaks as a function of time at a fixed applied potential of $E = 0.3 $~V. The ratio of trion to exciton decreases with time, which causes the superpeak to shift from between the two peaks at $1$~ps to entirely atop of the exciton peak at $25$~ps. The trion peak generally blueshifts over time whereas the exciton remains fairly constant, resulting in a blueshift of the superpeak, as originally observed in Fig~\ref{fig:initial_spectra}.}
\end{figure}

From this, one can address all aspects of the dynamical shift of Fig.~\ref{fig:initial_spectra}(right). For the $0.25$~V and $0.30$~V spectra, where the trion and exciton begin with similar intensity, the pronounced blueshift is mainly due to this change in relative heights. Indeed, both spectra converge to the high-potential/low-doping superpeak positions at long delay times. Additionally, the trion and exciton positions move with changing doping levels: the exciton is predicted to blueshift with increasing carrier density, while the trion is predicted to redshift. The shift in the underlying peaks thus explains the smaller movement of the superpeak at the highest and lowest doping levels considered, where the signal is dominated by a trion or an exciton, respectively. In general, the total shift has a contribution from both the changing ratio of peak heights and the changing ratio of BGR and BES. We summarize these effects in Fig.~\ref{fig:explanation} for the particular case of the $0.30$~V TA spectrum. 

This simple physical picture also explains why the superpeak redshifts before $1$~ps at intermediate voltages: the doping level is approximately constant (so the exciton and trion positions are fixed) but the formation of quasiparticles from free carriers has not completed.\cite{Ceballos2016, Cunningham2017} While the strongly enhanced absorption characteristic of free carriers is visible, the ratio of excitons to trions has not reached equilibrium. The superpeak redshifts because the trion intensity grows and saturates with time.

We arrive at a point where the MND~theory of Ref.~\onlinecite{Chang2019a} is in excellent agreement with our experimental steady-state absorbance and TA~data. We believe the explanation developed here also applies to the data in Refs.~\onlinecite{Sie2017,Bera2021} before the onset of the Mott transition. Here, the use of this minimal model to disentangle the spectral signals as a function of potential and time leaves one observation unexplained. For the spectra at the extreme potentials, we claimed that the peak shift is explained by the doping dependence of the exciton or trion signal. Applying the model independently to each time point resulted in deconvolved exciton and trion peak positions that quantitatively agree with the experimental data as a function of time (compare Appendix Fig.~\ref{fig:si3} with Fig.~\ref{fig:initial_spectra}(right)). Yet, both exciton and trion peaks net-blueshift in time. This observation fundamentally contradicts the logic that the exciton should blueshift with \textit{increasing} carrier density. Since carrier density decreases with time, the current model cannot accommodate this observation without further extension. We emphasize that, had we applied a data-driven technique, such as SMCR,\cite{Ruckebusch2020, DeJuan2007} without carefully including the physical constraints placed by the underlying Hamiltonian, the optimization procedure would likely have subsumed this subtle effect into a better overall fit; such an approach runs the risk of obtaining an aesthetically pleasing but physically misleading result. 

Several factors can cause the exciton peak to blueshift over time. First, temperature tunes the absorption spectrum and especially the trion signal.\cite{Zhang2014, Christopher2017} Importantly, we neglect nonequilibrium energy relaxation after photon absorption, which leads to changing temperatures and even phonon bottlenecks.\cite{Chi2020,Wang2021h} Ultimately, the timescale required for the lattice to cool the electronic degrees of freedom in comparison to our TA~timescales remains unclear. Second, BGR and BES changes may depend on the identity of the quasiparticles in the system at a particular time.\cite{Cunningham2017} For example, biexcitons,\cite{Li2018c, Trovatello2022, Muir2022} the next-highest-order quasiparticle, are not considered in our model. Finally, while the MND theory assumes holes with infinite mass in the exciton and trion, holes in the real system have an associated momentum.\cite{Qiu2015a} What additional physics is still required to constitute a minimal model therefore remains an open question.

Thus, we have interrogated the ability of trion formation as described by the MND Hamiltonian to explain the salient features in our linear and TA spectra. While a fully \textit{ab initio} treatment of the problem remains beyond the scope of existing theory, we have developed an algorithm to exploit the behavior of the spectra obtained for this model as a function of conduction band population to constrain a minimal parametrization of the data. Our physically meaningful approach captures the evolution of the trion and exciton peaks as a function of applied potential and thus offers a means to critically assess the ability of this model to describe the full range of observed phenomenology. Our results demonstrate that our simple approach is sufficiently robust to interpret, assign, and analyze the spectral features in linear and time-resolved TA spectra of doped ML-TMDs. Importantly, our analysis explains how complex, time-, doping-, and fluence-dependent spectral shifts previously attributed solely to the A-exciton instead arise from the convolution of the emergence and shift of the trion peak and the shift in the exciton peak. The appearance of the trion at room temperature, under working conditions, suggests that a model of TMD photophysics restricted to BGR and BES is insufficient to capture both static and dynamic spectral signatures, and that the trion state must therefore be included in the minimal theoretical framework for spectroscopic measurements of TMD-containing systems.

\section{Acknowledgements}
This research was supported by the U.S.~Department of Energy, Office of Science, Office of Basic Energy Sciences, under Award DE–SC0021189 (J.B.S., R.A.), and under Award DE-SC0016137 (A.T.K., Y.R.F.). A.M.C.~acknowledges the start-up funds from the University of Colorado, Boulder. T.S.~and A.M.C.~thank Lachlan Lindoy and David Reichman for sharing their 2D~spectroscopy code published in Ref.~\onlinecite{Lindoy2022}.

\onecolumngrid
\vfill
\pagebreak
\twocolumngrid
\appendix*

\section{Computational Details}\label{app:compdeets}
\vspace{-4pt}
\subsection{Spectra Pre-processing}\label{app:preprocess}
\vspace{-10pt}

\begin{figure}[h]
\begin{center}
    \resizebox{.45\textwidth}{!}{\includegraphics{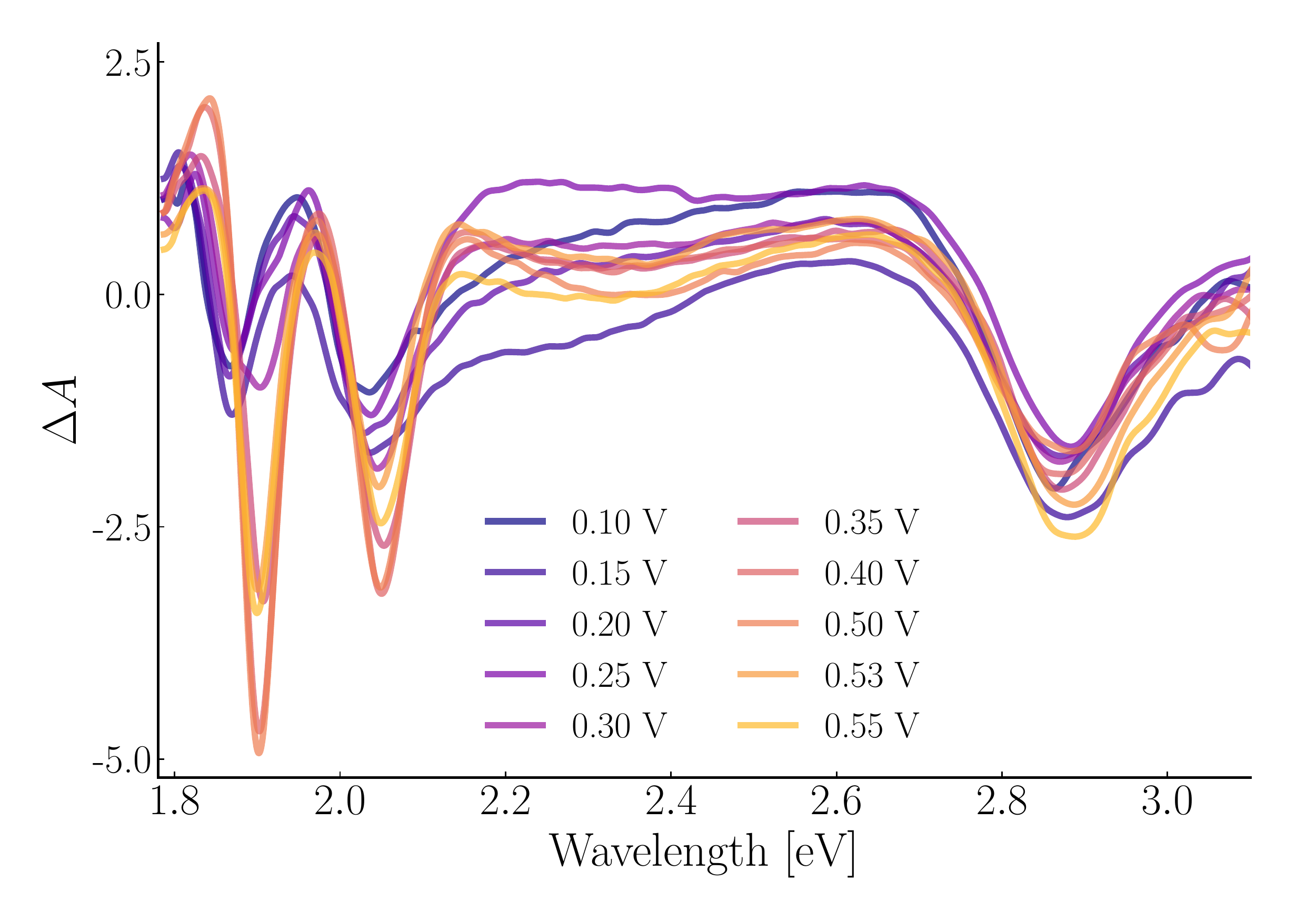}}\vspace{-2pt}
\end{center}
\vspace{-14pt}
\caption{\label{fig:sibar1} TA spectra at 1~ps seen in Fig.~\ref{fig:initial_spectra}, left, showing the full range of the measurement, for completeness. An additional trace at 0.10~V that was not used in the text is also shown. Its $\Delta A$ profile is roughly as expected, but the values are anomalously large.}
\end{figure}

In Fig.~\ref{fig:sibar1} we present the full TA spectra at 1~ps. Measurements at 0.10~V (shown), 0.00~V and 0.33~V (not shown) were also taken in the experiment, but these three traces were not stable across the delay window, and so they were omitted from the analysis. No smoothing is performed on the TA data.

Our steady state data are noisy in comparison to the TA spectra, so we smooth the experimental points using a Savitzky-Golay (SavGol) as distributed with the freely available $\mathrm{scipy.signal}$ library. The filter is chosen to have a window length of 17, and a polynomial order of 3. These are chosen conservatively to produce well-defined peak positions. We plot both the smooth and the data points on both relevant figures for transparency.

Both linear and TA spectra must be baselined for direct comparison with our model. In both cases we opt for a linear background subtraction. We do not account for the B-exciton around 2.05~eV. For these linear spectra, the background becomes linear between the A- and B-exciton features, so the overlap is small compared to the heights of the peaks. This may have some contribution to the poorer fit of the blue-side in the linear spectra. For the TA, one of the photoinduced features of the B-exciton will overlap with that of the A-exciton, but we have no good way of removing it at this time; we note the positive features (to the blue) of the B-exciton appear smaller in comparison. 

\begin{figure}[!h]
  \centering
  \begin{minipage}[c]{0.2\textwidth}
    \resizebox{1.0\textwidth}{!}{\includegraphics{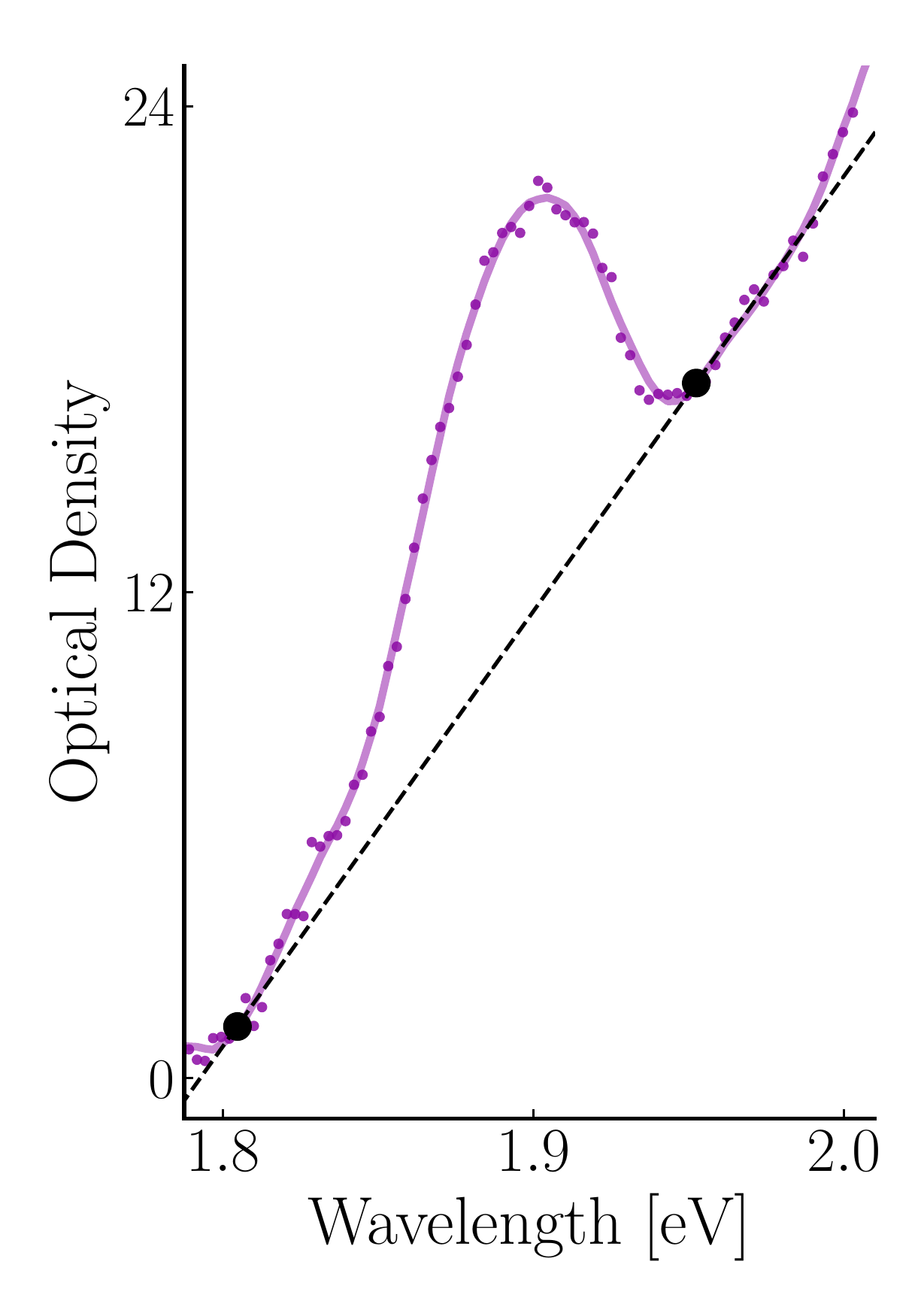}}
  \end{minipage}
  \begin{minipage}[c]{0.2\textwidth}
    \resizebox{1.0\textwidth}{!}{\includegraphics{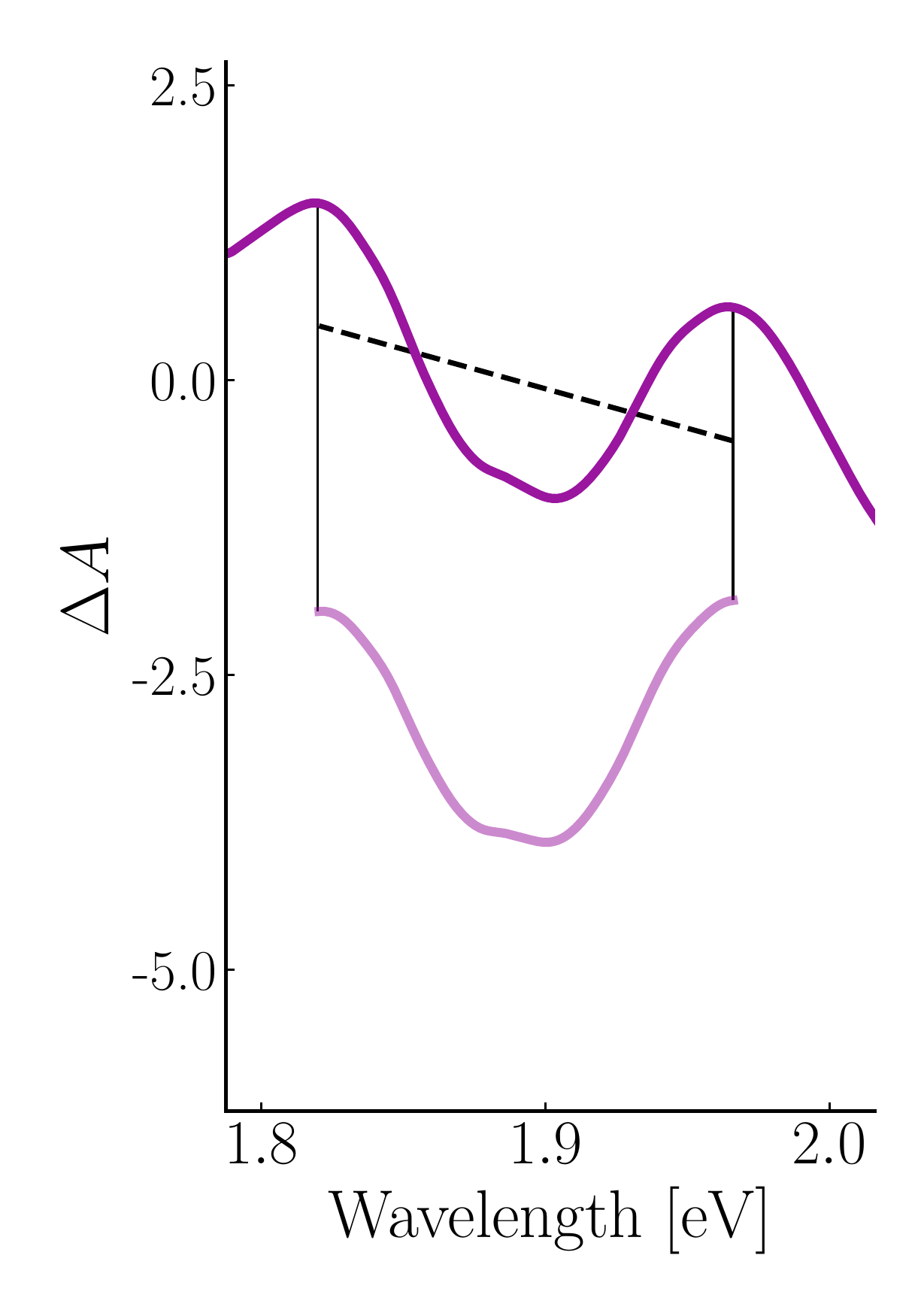}}
  \end{minipage}
  \vspace{-6pt}
  \caption{\label{fig:sibar2} Illustrative plots showing the background subtraction methodology. Both are for applied voltages of 0.3~V \textbf{Left}: Steady state spectrum, where the full black circles are the endpoints for the dashed black line. Subtracting this line results in the trace shown in Fig.~\ref{fig:steadystate}, right, of the main text. \textbf{Right}: TA spectrum at 1~ps, where the vertical black lines show the peak positions that delimit the linear fit. Subtracting this fit, shown as a dashed black line, results in the lighter purple curve which, after baselining, is the trace of Fig.~\ref{fig:TAfit1ps}.
  }
  \vspace{-6pt}
\end{figure}

In the case of the steady state data of Fig.~\ref{fig:steadystate}, we choose the lowest-gradient points either side of the 1.9~eV peak region and define a linear function between them that defines the background. An illustration of this is provided for the 0.3~V trace in Fig.~\ref{fig:sibar2}, left. This results in a very small negative region to one side of the peak, owing to the imprecise choice. The negative region is cropped. The position of the peak is fairly insensitive to moving the endpoints of this fit, at least for the data shown in this paper: there is enough of a linear region either side of the peak.

For the TA data the background subtraction is somewhat imperfect, for the reasons discussed in the main text. Here, the positive (photoinduced) features either side of the main bleach are identified, and a linear fit (not line) is performed between them. This is defined as the background. Again, we provide an illustration for the 0.3~V trace, in Fig.~\ref{fig:sibar2}, right. The gradient of this fit does not change hugely expanding the window to the red and blue respectively, as the traces are fairly symmetrical---in this case the curves are then cropped at the resulting maxima on each side, which may move slightly.

\vspace{-14pt}
\subsection{Optimization Procedure}
\label{app:optimization-procedure}
\vspace{-6pt}
Here we outline how the model parameters are extracted from the potential-dependent photoelectrochemical spectra at a particular time. We begin with a top-level summary of the method, and then we will provide the in-depth technical details and code.

At its most basic, the problem is one of resolution. For these data, the optical response saturates as the applied potential reaches its largest values. In the MND description, this saturation occurs because the superpeak has only exciton contribution, as the conduction band becomes fully depleted. Therefore, at this zero doping condition, we can identify the exciton peak position, height, and broadening. However, other values, such as the trion position at low doping (because there is no intensity), are challenging to extract by eye. Thus, to obtain a good choice of the model parameters that agree with the experimental data, we developed the following Monte Carlo-based fitting algorithm: 
\pagebreak
\begin{enumerate}
    \item For a given set of parameters consistent with the output of an MND model we obtain---after broadening with the (parametrically fixed) Gaussian---a particular manifold of curves, such as that displayed in Fig.~\ref{fig:steadystate}(d).
    \item Each experimental superpeak, having been isolated from the sloping background, can then be paired with its counterpart in this manifold that has the smallest point-wise deviation. 
    \item The sum of all the best-fit errors from the `global' set defines a total error. 
    \item We make a small change to the parameters, subject to the constraints established by the MND Hamiltonian, and repeat steps 1--3.
    \item We accept the change made in step~4 if it lowers the total error, else we reject it with a probability decreasing with error (Metropolis Monte Carlo).
\end{enumerate}
We iterate this procedure until the error converges.

\begin{figure}[h]
\begin{center}
    \resizebox{.3\textwidth}{!}{\includegraphics{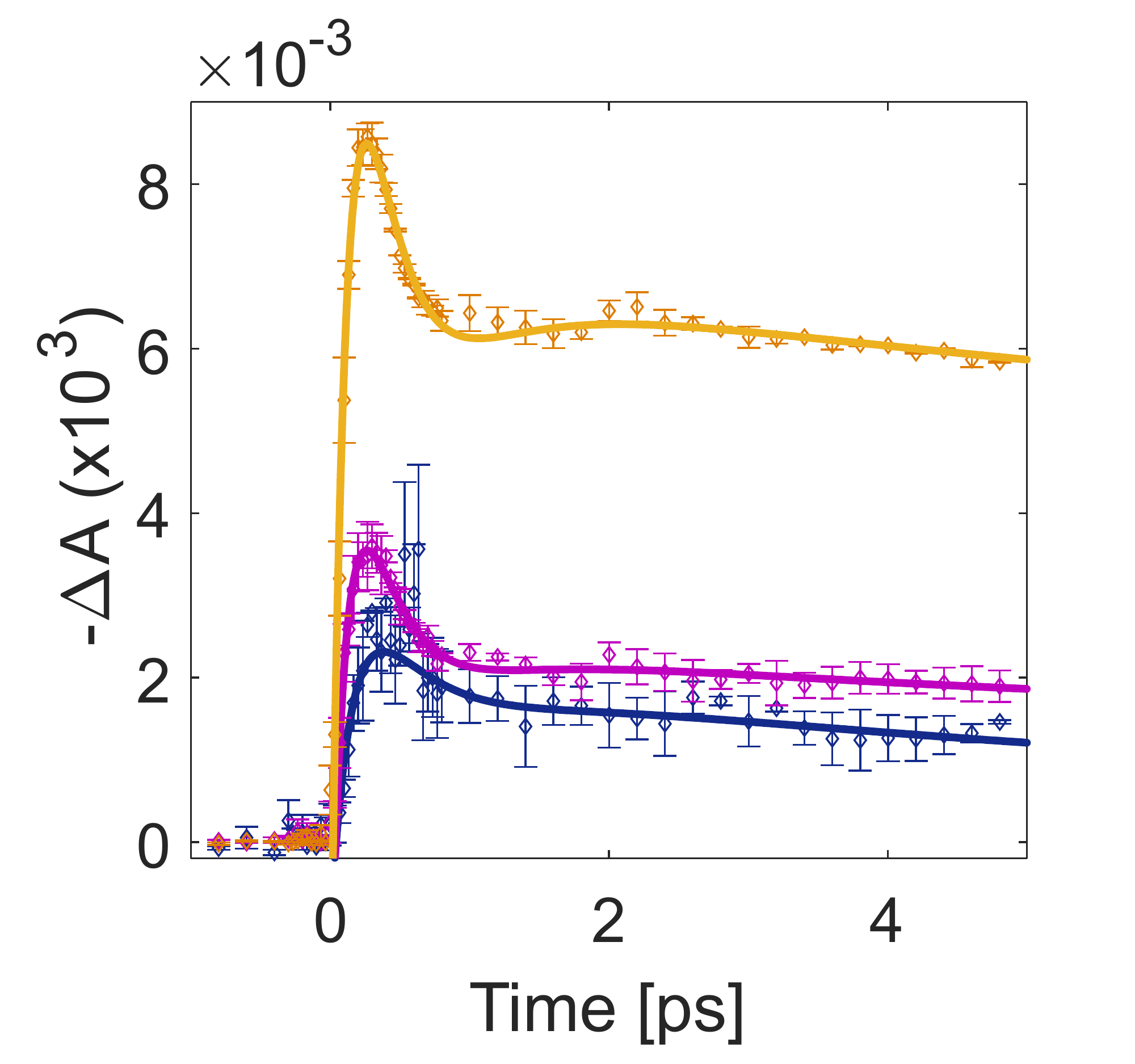}}\vspace{-2pt}
\end{center}
\vspace{-18pt}
\caption{\label{fig:sibar0.5} TA spectra at 0.55~V (yellow), 0.30~V (magenta), and 0.10~V (blue) at the A-exciton `superpeak' wavelength as a function of time. The short-lived decay component before 1~ps has been assigned as a non-equilibrium signature of free-carriers waiting to form excitons (and therefore trions). While the effect looks larger at higher potential, it should be noted that there is an accompanying shift (Fig.~\ref{fig:initial_spectra}, right) which we assign as a peak-splitting. As is discussed throughout the text superpeak parameters should not be directly interpreted.}
\end{figure}

At the more technical level, there are a few practical details that need to be addressed. Pseudoised code of the important functions is provided for clarity. First, as described in the main text, the TA spectra before 1~ps display a strong, short-lived `spike' feature. For completeness, we display time-series data showing this feature  at the A-exciton (superpeak) wavelength at low, medium, and high voltage in Fig.~\ref{fig:sibar0.5}. As this effect has been assigned to relaxing free-carriers,\cite{Ceballos2016, Cunningham2017} we focus on the post 1~ps regime where the interpretation of the fitted parameters is more straightforward, and their values at consecutive time points ought to be good guesses for one another.

\vspace{-14pt}
\subsubsection*{Steps 1--3}
\vspace{-10pt}

For a given pre-processed set of curves, we begin with an initial guess. When running on the full time series, this guess is the solution to the previous time step (for which a solution exists). For the steady state spectra, or for the first (1 ps) TA spectra, the guess is made by hand. For Fig.~\ref{fig:TAfit1ps}, this takes in the following parameters,

\begin{minted}[fontsize=\scriptsize]{python} 
pos_start_X = 1.899; pos_end_X = 1.909
pos_start_T = 1.876; pos_end_T = 1.865
heights_X = [1.1, 0.9, .7, 0.4, 0.15, 0.05, 0.00]
heights_T = [0.00, 0.01, 0.02, 0.06, 0.11, 0.17, 0.25]
width_start_X = 2.0E-3; width_end_X = 12.0E-3
width_start_T = 5E-3; width_end_T = width_start_T
\end{minted}

using them to produce curves like those of Figure~3 in Ref.~\onlinecite{Chang2019a}, i.e. curves that respect the constraints of the MND model. In particular, the position and width parameters define linear functions between the starting and ending doping levels. The heights are instead a quadratic fit to the list of points provided (uniform over the doping region); seven points were used to stabilise the fit against noise during the Monte Carlo loop (\textit{vide infra}), as when only three points were used the rejection rate of guesses was unacceptably high. Using a set of variable points was quicker to implement than finding the constraints required to make the quadratic fit monotonically decrease over the doping range. Fortunately, this less-robust approach allowed us to find the positive trion values discussed in the text.

Once these doping curves have been obtained, the manifold pictured in Fig.~\ref{fig:steadystate}, right, can be constructed. Specifically,

\begin{minted}[fontsize=\scriptsize]{python} 
instr = 6E-2 * gaussian(x_axis, (x_upper + x_lower) 
        / 2, 1.1 * 5.0E-2)  # measurement broadening
manifold = np.array([np.zeros_like(x_axis) for E in E_axis])
for i, E in enumerate(E_axis):
    X_peak = height_X_f(E) * 
        lorentz(x_axis, pos_X_f(E), width_X_f(E), E)
    T_peak = height_T_f(E) * 
        lorentz(x_axis, pos_T_f(E), width_T_f(E), E)
    XT_peak = X_peak + T_peak
    peak = np.convolve(instr, XT_peak, "same")
    manifold[i] = peak
\end{minted}

The doping window ($\mathrm{E}\_\mathrm{axis}$) is always chosen to be from 0--30~meV, as this ought to provide enough range in the first instance. However, as noted, changing the parameters can incorporate a stretching of the doping range which means that---particularly for the TA spectra---it should not be directly interpreted as matching that in Ref.~\onlinecite{Chang2019a}. Instead, a feature like the crossover point between trion and exciton should be matched between the original model and the parameterized curves, if such a comparison is desired. The `gaussian' function has unit height, and the `lorentz' function is defined as
\begin{minted}[fontsize=\scriptsize]{python} 
def lorentz(p, p_0, w, E=0.03):
    w_p = 2. * w / (1. + np.exp(-(100 * E / 0.03) * (p - p_0)))
    x = (p - p_0) / (w_p / 2)
    return 1. / (1. + np.power(x, 2))
\end{minted}
where the asymmetry parameters is set to $100$ to achieve agreement with the plots in Ref.~\onlinecite{Chang2019a}. This value is never changed.

Each of the experimental spectra can then be compared against the manifold to find their best fit, subject to the condition that increasing experimental voltage decreases the doping level from the previous value. Further, the fit cannot move more than one quarter of the doping range in one go, as this would not line up with the experimental observation and is just done for sanity. Roughly speaking, the algorithm cannot jump around on the doping axis and must respect the voltage ordering. Additionally, the best fit choice downweights the error away from the central region by around 50\% using a Gaussian mask: this is because the tails were never fitted particularly well, as discussed in the main text. The mask was chosen by reducing the value (Gaussian width) until the fit produced peak positions that were in agreement for the pure exciton. In code,

\begin{minted}[fontsize=\scriptsize]{python} 
def choice_E_fit(*args, **kwargs):
    ...
    minerr = len(manifold)  # we start the scan from the right
    maxstep = int(len(manifold)/4) # we won't go more in one step
    besterrs = []
    for i, voltage in enumerate(voltages):
        else:
            x_L = peak_xaxes[i][0]
            x_R = peak_xaxes[i][-1]
            cfit = np.copy(peaks[i] / sq)  #normalization factor
            gfilter = gaussian(x_axis[x_L:x_R], 
                    x_axis[int(x_L/2+x_R/2)], 
                    (x_axis[x_R]+x_axis[x_L]) / 30)
            # 30 is 50% at the wings, 50 is 10% at the wings
            errors = np.zeros_like(E_axis)
            for k, theory in enumerate(manifold):
                tfit = theory[x_L:x_R]
                diff = cfit - tfit
                diff *= gfilter
                errors[k] = (np.linalg.norm(diff))
            # we don't want to scan too far
            steplim = max(0, minerr-maxstep)  # need to clamp it at 0
            # dont overshoot the previous fit value
            steplim += np.argmin(errors[steplim:max(1, minerr)])
            # max is to stop it crashing due to empty list
            choice_E = E_axis[steplim]
            choice_Es.append(choice_E)
            besterrs.append(errors[steplim])  # we need this for MC
    choice_Es.reverse()  # it will be low to high this way
    return choice_Es, besterrs
\end{minted}

\vspace{-14pt}
\subsubsection*{Steps 4 and 5}
\vspace{-10pt}
Now a single iteration has been defined, the Monte Carlo loop may be performed. Steps in parameter space are defined in a way that respects the MND model. From the reference parameters (previous accepted value) the small changes are chosen by

\begin{minted}[fontsize=\scriptsize]{python} 
def make_delta(reference, loop_params, min_BE=0.005, min_shift=0.000):
    delta_params = copy.copy(reference)
    # keep the beginning exciton position fixed
    delta_params[0] = 0
    # its end-point can vary more, but must be bluer
    # the exciton MUST shift with potential. 
    # This kwarg is to stop is being static, which can happen.
    delta_params[1] = max((loop_params[0]+delta_params[0]-
                                    loop_params[1]+min_shift), 
                                    np.random.normal(0, 5E-3))
    # the trion is harder to fit
    # it must be more red
    # however they might go ontop of each other, 
    give a minimum binding energy: min_BE kwarg
    delta_params[2] = min((loop_params[0]+delta_params[0]-
                                    loop_params[2]-min_BE), 
                                np.random.normal(0, 1E-2))
    # the trion must always be more red than the exciton, 
    # good enough to make the start lower, since it gets more red
    delta_params[3] = min((loop_params[2]+delta_params[2]-
                                            loop_params[3]), 
                                    np.random.normal(0, 1E-2))
    # now heights.
    # exciton starts at 1 by normalization
    # quadratic fit musn't have a turning point, no minimum...
    # TODO
    hX = delta_params[4]
    hT = delta_params[5]
    hX[0] = max((0.95-loop_params[4][0]), 
            min(1.05-loop_params[4][0], 
                np.random.normal(0, 0.2)))
    # first max: don't let it drop too much from 1.0
    # second min: the data are normalized
    hT[0] = 0  # no trion at the start
    # we know its close to zero, right now first argument is!
    # the trion never gets above about 0.6 units
    for i in (1, 2, 3, 4, 5, 6):
        hX[i] = max(-loop_params[4][i],
                    min(loop_params[4][i - 1] + hX[i - 1] - 
                                        loop_params[4][i], 
                                np.random.normal(0, 0.2)))
        # first max/min to stop extreme new_params,
        # second min to stop overshoot previous value
        hT[i] = max(-loop_params[5][i], 
                min(0.6-loop_params[5][i],
                    max(loop_params[5][i - 1] + hT[i - 1] -
                                        loop_params[5][i], 
                                np.random.normal(0, 0.05))))
        # difference here is trion is growing, so max not min
    # now the widths
    delta_params[6] = max(-loop_params[6] + 1E-4,
                        np.random.normal(0, 5.0E-4))  
    # widths cannot be too small
    # has to end wider than it started
    delta_params[7] = max(-loop_params[7] + loop_params[6] + 
                                            delta_params[6], 
                                np.random.normal(0, 3.0E-3))  
    # there's a lot of possibility here
    # this needs to be close to where the exciton starts
    delta_params[8] = max(loop_params[6] + delta_params[6] -
                                            loop_params[8], 
                                np.random.normal(0, 1.0E-4))
    delta_params[9] = delta_params[8]  # fixed same by theory
    
    return delta_params
\end{minted}
which is sufficiently general that it did not need to be changed. The Monte Carlo loop itself is done in a staged way, where many small steps are compared before attempting a Metropolis test.

\begin{minted}[fontsize=\scriptsize]{python} 
def run_MC(*args, **kwargs):
    ...
    """tests on the first spectrum suggested that you can
    find a rare improvement in a sample of 3000 or 4000,
    so I would make this the inner loop"""
    choice_es, besterrs = choice_E_fit(params)  # initial guess
    err_hand = np.sum(besterrs)
    print(f"the overall error is {err_hand}")
    best_overall = [params, choice_es, err_hand]
    loop_params = copy.copy(params)
    
    # now we need the loop that will vary the parameters
    def inner_search(new_params):
        fit_funcs = make_regime(*new_params)  # turn into curves
        choice_es, besterrs = choice_E_fit(new_params)
        toterr = np.sum(besterrs)
        return new_params, choice_es, toterr
        
        
    np.random.seed(2022)  # year of our Lord.
    for j in range(outer_loop):
        # the outer loop performs the MC
        print(f"------OUTER LOOP IS NOW AT {j}-------")
        results = [0 for i in range(inner_loop)]
    
        # SERIAL version, in the real code this is parallel
        t_inner = perf_counter()
        for k in range(inner_loop):
            results = inner_search(add_params(loop_params, 
                make_delta(reference, loop_params)))
        # add_params does what it sounds like
    
        best_in_show = min(results, key=lambda t: t[2])
        if best_in_show[2] < best_overall[2]:
            print(f"OUTER LOOP lowers error to {best_in_show[2]}")
            accept = 1
        else:
            prob_metro = np.exp(
                    (best_overall[2] - best_in_show[2]) / temper)
            # choose temper to get sensible reject rate
            if metro>0: print(f"{best_overall[2]} -> 
                        {best_in_show[2]} => {100*prob_metro:.0f}%")
            if np.random.uniform() < prob_metro and metro>0:
                accept = 1
                print(f"ACCEPTS worse result in OUTER LOOP")
            else:
                if metro>0: 
                    print(f"REJECTS worse result in OUTER LOOP")
                accept = 0
    
        if accept == 1:
            loop_params = best_in_show[0]
            best_overall = best_in_show
            with open(pkl_end, "wb") as f:
                pickle.dump(best_overall, f)  # save to disk
    
    print("------ALL OUTER LOOPS COMPLETE-------")
\end{minted}
This is done because the random steps in parameter space are chosen brute force, not in some downhill-seeking way. Backpropagation would probably be a far more efficient way of choosing steps, but this code was not found to be expensive to run locally. The files saved to disk can then be used as the guess for the next time step, and for the analysis presented in the main text.

\vspace{-14pt}
\subsection{TA data results}\label{sec:TAresults}
\begin{figure}[!h]
\vspace{-12pt}
\begin{center}
    \resizebox{.4\textwidth}{!}{\includegraphics{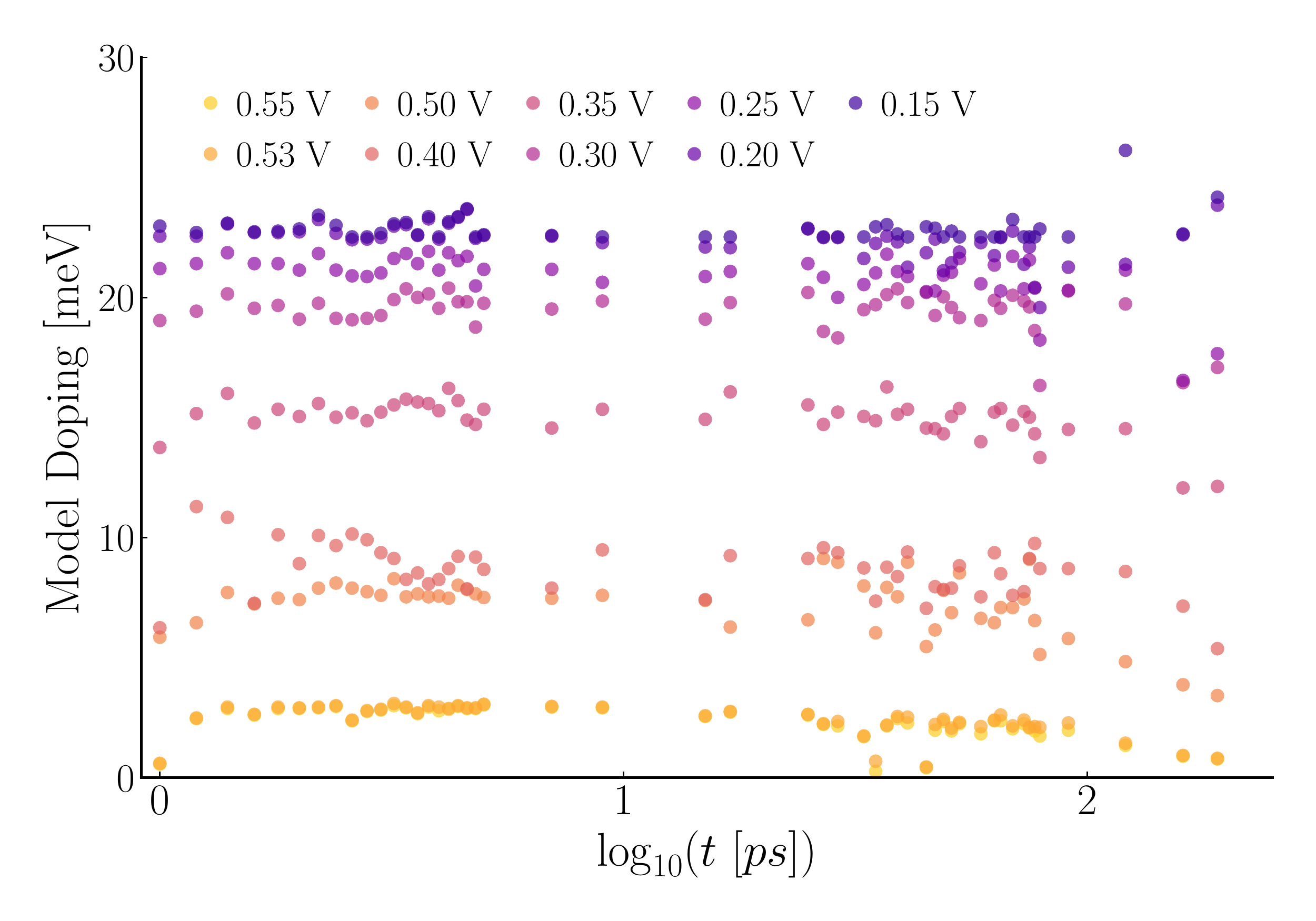}}\vspace{-2pt}
\end{center}
\vspace{-20pt}
\caption{\label{fig:si1} Doping level assigned to each experimental trace (voltage) as a function of time. Since we contend that the conduction band density should be mostly due to the background carriers from the applied potential, we expect our fitting algorithm to assign a roughly constant doping level with time. The ordering of the values is enforced by the algorithm, e.g. the 0.15~V trace must be assigned a higher value than the 0.2~V trace.}
\vspace{-6pt}
\end{figure}

\begin{figure}[!h]
\begin{center}
    \resizebox{.4\textwidth}{!}{\includegraphics{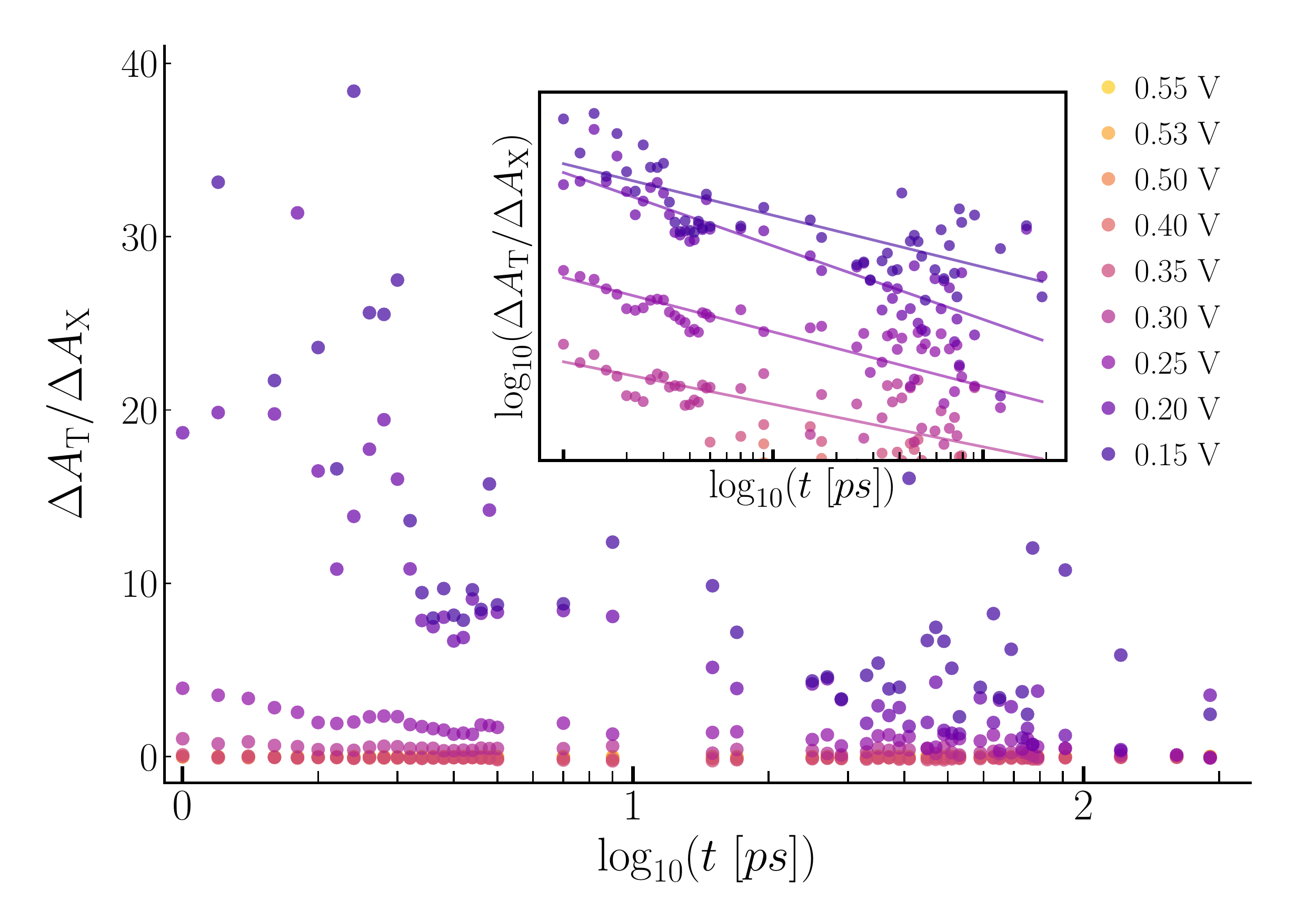}}\vspace{-2pt}
\end{center}
\vspace{-20pt}
\caption{\label{fig:si2} Fig.~\ref{fig:TAresults}, left, but taking the ratio of bottom panel to top panel. The inset shows the result of taking the logarithm of the four traces with entries greater than one (trion dominant). Approximate single-exponential behaviour is observed.}
\vspace{-6pt}
\end{figure}

\begin{figure}[!h]
\begin{center}
    \resizebox{.4\textwidth}{!}{\includegraphics{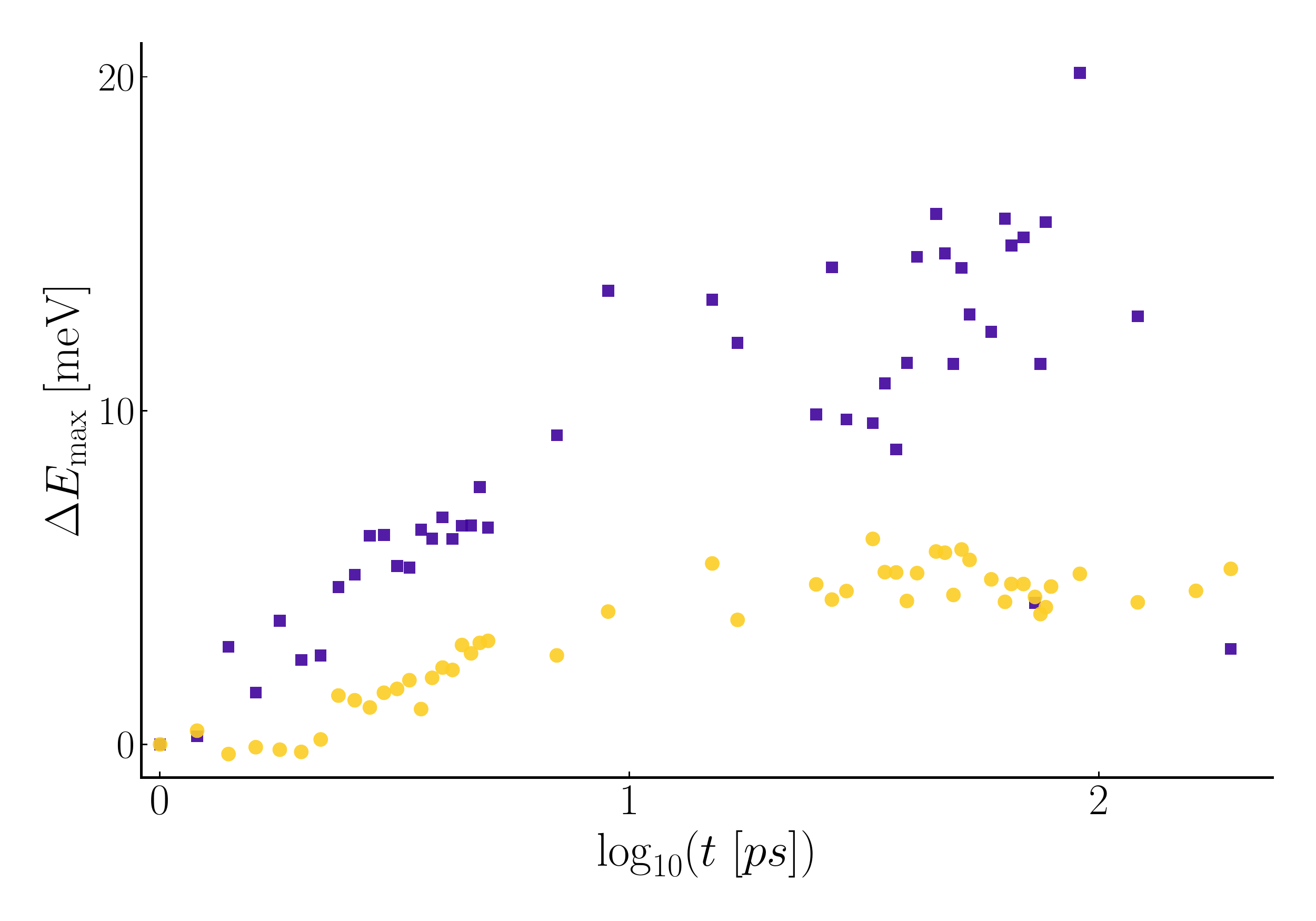}}\vspace{-2pt}
\end{center}
\vspace{-20pt}
\caption{\label{fig:si3} Shift of the peaks corresponding to extreme doping levels (applied voltage) as predicted by the model. Yellow circles are the shift of the 0 meV exciton, blue squares are the shift of the 23 meV trion, which is the (approximate) assignment of the 0.15~V experimental trace, as shown in Fig.~\ref{fig:si1}.}
\vspace{-14pt}
\end{figure}

\vspace{-10pt}
Figure~\ref{fig:si1} is provided to support the claim that the Monte Carlo method is achieving meaningful fits across time. Each fit is independent---only using the previous result as an initial guess---and can in principle give completely different results to its neighbours if the fit is poor. The stability of the doping level, and the fact that Figs.~\ref{fig:si2}~and~\ref{fig:si3} show a continuousness across time is supporting evidence, along with the visual goodness of fit, that the algorithm is performing adequately well.

Gaps in the time series around 1~ps are due to experimental anomalies where at least one of the traces displays erroneous intensity behaviour (e.g. the 0.3~V trace is suddenly twice as strong) or the absence of any well-defined peaks. Since these are confined usually to only one or two traces at a time, and not seen at the next time point, these are diagnosed as temporary interruptions to the data collection process. To be consistent, the entire set of spectra are omitted at each of these points, which were all confirmed by manual inspection.

\vspace{-14pt}
\subsection{2D spectroscopy MND code}
\vspace{-10pt}
As the code\cite{Lindoy2022} is a direct extension of Ref.~\onlinecite{Chang2019a}, it is already programmed with our default values. We ran for 40~fs, and moved the exciton peak to the appropriate energy value (1.9~eV rather than 2.0~eV). The position of the peak is entirely arbitrary (sets the energy zero) for this Hamiltonian. There is no bath, only relaxation to the ground state.

\vfill\pagebreak
~~~
\pagebreak
\subsection*{References}
\vspace{-14pt}
\bibliography{Postdoc-TMDC}

\end{document}